\documentclass[10pt,x11names,twocolumn]{article}

\usepackage[T1]{fontenc}
\usepackage[latin1]{inputenc}

\usepackage{hyperref}

\usepackage{fancyhdr,cancel}

\usepackage{listings}
\usepackage{enumitem,multicol,xcolor,comment}

\usepackage{times}

\usepackage[hang,small,labelfont=bf,up,textfont=it,up]{caption}
\usepackage{subcaption}
\usepackage{amsmath,amsfonts,amsthm,amssymb,bm,dsfont,booktabs}	

\usepackage{setspace}
\usepackage{enumitem}
\setlist{nolistsep}

\usepackage{graphicx}

\usepackage{adjustbox}

\usepackage[letterpaper,top=2.75cm,bottom=2.75cm,left=2cm,right=2cm]{geometry}

\usepackage{dutchcal}
\usepackage{centernot}

\usepackage{numprint}
\npthousandsep{,}\npthousandthpartsep{}\npdecimalsign{.}

\newcommand{\abs}[1]{\left\vert #1 \right\vert}
\renewcommand{\vec}[1]{\boldsymbol{#1}}
\newcommand{\mvec}[1]{\mathbf{#1}}
\newcommand{\dplus}{d_+}
\newcommand{\dminus}{d_-}

\newcommand{\frob}{\text{F}}
\newcommand{\aip}{\text{AIP}}
\newcommand{\ipa}{\text{IPA}}

\usepackage{thmtools,url}

\newcommand{\titledoc}{Link prediction in dynamic networks \\ using random dot product graphs}
\newcommand{\titleshort}{Link prediction in dynamic networks using random dot product graphs}

\usepackage{csquotes} 
\usepackage{natbib}

\usepackage{sectsty}
\partfont{\fontsize{17}{17}\selectfont}

\usepackage{thmtools}
\newtheorem{definition}{Definition}

\usepackage{authblk}

\usepackage[ruled,linesnumbered]{algorithm2e}
\SetKwInput{KwInput}{Input}
\hypersetup{colorlinks,linkcolor={blue},citecolor={blue},urlcolor={blue}}  
\numberwithin{equation}{section}

\usepackage{mathtools}
\mathtoolsset{showonlyrefs}

\providecommand{\keywords}[1]{\textbf{\textit{Keywords ---}} #1}

\author[1]{Francesco Sanna Passino}
\author[2]{Anna S. Bertiger}
\author[2]{Joshua C. Neil\thanks{The author is currently at Securonix Threat Labs, Securonix Inc., Addison (TX). This work was completed when the author was at Microsoft 365 Defender, Microsoft Corporation, Redmond (WA).}}
\author[1]{Nicholas A. Heard}

\affil[1]{Department of Mathematics, Imperial College London}
\affil[2]{Microsoft 365 Defender, Microsoft Corporation, Redmond (WA)}

\date{}

\title{\huge\textbf{\titledoc}}
	
\begin{document}


\maketitle



\begin{abstract}
The problem of predicting links in large networks is an important task in a variety of practical applications, including social sciences, biology and computer security. In this paper, statistical techniques for link prediction based on the popular random dot product graph model are carefully presented, analysed and extended to dynamic settings. Motivated by a practical application in cyber-security, this paper demonstrates that random dot product graphs not only represent a powerful tool for inferring differences between multiple networks, but are also efficient for prediction purposes and for understanding the temporal evolution of the network. The probabilities of links are obtained by fusing information at 
two stages: spectral methods provide estimates of latent positions for each node, and time series models are used to 
capture temporal dynamics.
In this way, traditional link prediction methods, usually based on decompositions of the entire network adjacency matrix, are extended using 
temporal information. 
The methods presented in this article are applied to a number of simulated and real-world graphs, showing promising results. 
\end{abstract}

\vspace{1em}
\keywords{adjacency spectral embedding, dynamic networks, link prediction, random dot product graph.}

\section{Introduction} \label{sec:introduction}

Link prediction is defined as the task of predicting the presence of an edge between two nodes in a network, based on latent characteristics of the graph \citep{liben}. 
The problem 
has been widely studied in the literature 
\citep{lu,menon}, and has relevant applications in a variety of different fields. In this paper, the discussion 
about link prediction 
is motivated by applications in cyber-security and computer network monitoring \citep{Jeske18}.  
The ability to correctly predict and associate anomaly scores with the connections in a network is valuable for the cyber-defence of enterprises. In cyber settings, adversaries may introduce changes in the structure of 
an enterprise network in the course of their attack. Therefore, predicting links in order to identify significant deviations in expected behaviour could lead to the detection of an otherwise extremely damaging network breach. 
In particular, it is necessary to correctly score {\it new links} \citep{Metelli19}, representing previously unobserved connections. The task is particularly important since it is common to observe malicious activity associated with new links \citep{Neil13}, and it is therefore crucial to understand the normal process of link formation in order to detect a cyber-attack. 

In this article, it is assumed that snapshots of a dynamic network are observed at discrete time points $t=1,\dots,T$, obtaining a sequence of graphs $\mathbb G_t=(V,E_t)$. The set $V$ represents the set of nodes, which is invariant over time, 
whereas the set $E_t$ is a time dependent edge set, where $(i,j)\in E_t$ for $i,j\in V$, if $i$ connected to $j$ at least once during the time period $(t-1,t]$. Each snapshot of the graph can be characterised by the adjacency matrix $\mvec A_t\in\{0,1\}^{n\times n}$, where $n=\abs{V}$ and for $1\leq i,j\leq n$, $A_{ijt}=\mathds 1_{E_t}\{(i,j)\}$, such that $A_{ijt}=1$ if a link between the nodes $i$ and $j$ exists in $(t-1,t]$, and $A_{ijt}=0$ otherwise. The graph is said to be undirected if $(i,j)\in E_t \iff (j,i) \in E_t$, implying that $\mvec A_t$ is symmetric;
otherwise, the graph is said to be directed. It will be assumed that the graph has no self-edges, implying $\mvec A_t$ is a hollow matrix. Similarly, bipartite graphs $\mathbb G_t=(V_1,V_2,E_t)$ can be represented using two node sets $V_1$ and $V_2$, and rectangular adjacency matrices $\mvec A_t\in\{0,1\}^{n_1\times n_2}$, $n_1=\abs{V_1}, n_2=\abs{V_2}$, where $A_{ijt}=1$ if $i\in V_1$ connects to $j\in V_2$ in $(t-1,t]$.

This paper discusses methods for \textit{temporal link prediction} \citep{Dunlavy11}: given a sequence of adjacency matrices $\mvec A_1,\dots,\mvec A_T$ observed over time, the main objective is to reliably predict $\mvec A_{T+1}$.
In this article, temporal link prediction techniques based on {\it random dot product graphs} \citep[RDPG,][]{Young07} are discussed and compared. RDPGs are a class of {\it latent position models} \citep{Hoff02}, and have been extensively studied because of their analytical tractability \citep{Athreya18}. 
Each node $i$ is given a latent position $\vec x_i$ in a $d$-dimensional latent space $\mathbb X$ such that $\vec x^\top\vec x^\prime\in[0,1]\ \forall\ \vec x,\vec x^\prime\in\mathbb X$. The edges between pairs of nodes are generated independently, with probability of a link between nodes $i$ and $j$ obtained through the inner product $\mathbb P(A_{ij}=1)=\vec x_i^\top\vec x_j$. In matrix notation, the latent position can be arranged in a $n\times d$ matrix $\mvec X=[\vec x_1,\dots,\vec x_n]^\top\in\mathbb X^n$, and the expected value of a single realised adjacency matrix $\mvec A$ 
is expressed as $\mathbb E(\mvec A)=\mvec X\mvec X^\top$.

Random dot product graph models and spectral embedding methods are often the first step in the analysis of a graph, because of their simplicity and ease of implementation, since intensive hyperparameter tuning is not required. RDPGs are extensively applied in neuroscience \citep[see, for example,][]{Priebe17}. Furthermore, they have appealing theoretical statistical properties in terms of consistency of the estimated latent positions. Therefore, it is of interest to understand their performance for link prediction purposes. 

RDPGs models for multiple heterogeneous graphs on the same node set have recently been proposed in the literature, but these models have not been formally extended to a dynamic setting for link prediction purposes. Early examples discuss methods for clustering and community detection with multiple graphs \citep{Tang09, Shiga10, Dong14}. More recently, the focus has been on testing for differences in brain connectivity networks \citep{Arroyo17,Ginestet17,Durante18,Kim19}. \cite{Levin17} propose an \textit{omnibus} embedding in which the different graphs are jointly embedded into a common latent space, providing distinct representations for each graph and for each node. \cite{Wang17} propose the \textit{multiple random eigen graph} (MREG) model, where a common set of $d$-dimensional latent features $\mvec X$ is shared between the graphs, and the inner product between the latent positions is weighted differently across the networks, obtaining $\mathbb E(\mvec A_t)=\mvec X\mvec R_t\mvec X^\top$, where $\mvec R_t$ is a $d\times d$ diagonal matrix. 
\cite{Nielsen18} propose the \textit{multiple random dot product graph} (multi-RDPG), which more naturally extends the RDPG to the multi-graph setting. Their formulation is similar to the MREG of \cite{Wang17}, but $\mvec X$ is modelled as an orthogonal matrix, and $\mvec R_t$ is constrained to be positive semi-definite.
The model is further extended in \textit{common subspace independent edge} (COSIE) graphs \citep{Arroyo19}, in which $\mvec R_t$ does not need to be a diagonal matrix. 
More recently, \cite{Jones21} proposed the Unfolded Adjacency Spectral Embedding (UASE) for the multilayer random dot product graph (MRDPG), which is also applied to a link prediction example within a cyber-security context.
In this work, existing methods for RDPG-based inference, for example omnibus embeddings \citep{Levin17} and COSIE graphs \citep{Arroyo19}, will be analysed for link prediction purposes, and compared to standard spectral embedding techniques.

The main contribution of this work is to adapt the existing methods for multiple RDPG graph inference for temporal link prediction. 
Furthermore, 
this article proposes methods to combine the information obtained via spectral methods with time series models, to capture the temporal dynamics of 
the observed graphs. The proposed methodologies will be extensively compared on real world and simulated networks. 
It will be shown that this approach significantly improves the predictive performance of multiple RDPG models, especially when the network presents a seasonal or temporal evolution. 
Overall, this article provides insights into the predictive capability of random dot product graphs, and gives guidelines for practitioners on the optimal choice of the embedding for temporal link prediction. 

Importantly, the strategies for combination of individual embeddings, and their time series extensions, 
can in principle be applied to \textit{any} embedding method for static graphs, despite the main focus on RDPGs of this article. 
This article primarily focuses on RDPGs because of the wide variety of embedding techniques which have been suggested in the literature under this model, but that so far have not been compared for link prediction purposes. 

The article is organised as follows: Section~\ref{litrev} discusses related literature around link prediction, and Section~\ref{grdpg_sec} introduces background on the generalised random dot product graph and adjacency spectral embeddings, the main statistical tools used in this paper. 
Methods for link prediction based on random dot product graphs are discussed in Section~\ref{dynamic_pred}. 
Section~\ref{ts_extension} presents techniques to improve the predictive performance of the RDPG models, based on time series models. Results and applications on simulated and real world networks are finally discussed in Section~\ref{results}.

\section{Related literature} \label{litrev}

Many other models other than RDPGs have been proposed in the literature for link prediction. 
Traditionally, the temporal link prediction task is tackled using tensor decompositions \citep{Dunlavy11}. Dynamic models have also been proposed in the literature of Poisson matrix factorisation and recommender systems \citep{charlin,Hosseini18}, and extended to Bayesian tensor decompositions \citep{Schein15}. 
In general, including time has been shown to significantly improve the predictive performance in a variety of model settings, for example 
stochastic blockmodels \citep{Ishiguro10,Xu14,Xing10}.
More generic latent feature models for dynamic networks have also been extensively discussed in the 
literature 
\citep{Sarkar06,Krivitsky14,Sewell15}.

Latent features are usually obtained via matrix factorisation, considering constant and time-varying components within the decomposition \citep{Deng16,Yu17,Yu17_2}. Usually, a Markov assumption is placed on the evolution of the latent positions \citep{Zhu16,Chen18}. \cite{Gao11} propose to combine node features into a temporal link prediction framework based on matrix factorisation. Nonparametric \citep{Sarkar14,Durante14} and deep learning \citep{Li14} approaches have also been considered. 

More recently, advancements have been made in the application of deep learning to graph-valued data. 
In particular, deep learning methods on graphs are classified by \cite{Zhang20} into five categories: graph recurrent neural networks, graph convolutional networks, graph autoencoders, graph reinforcement learning, graph adversarial methods. 
A comprehensive survey of existing static network embedding methods, including deep learning techniques, is provided in \cite{Cai18}. 
Commonly used static embedding methods in machine learning are DeepWalk \citep{Perozzi14}, SDNE \citep{Wang16}, node2vec \citep{Grover16}, GraphSAGE \citep{Hamilton17}, graph convolutional networks \citep[GCN,][]{Kipf17}, and Watch Your Step with Graph Attention \citep[WYS-GA,][]{Haija18}. 
Many of such methods could be unified under a matrix factorisation framework \citep{Qiu18}. A systematic comparison of some of the aforementioned methodologies for unsupervised network embedding is provided in \cite{Khosla21}.
Methodologies have also been recently proposed in the dynamic network setting, within the context of representation learning \citep[for example,][]{Nguyen18,Kumar19,Liu19,Qu20}, and deep generative models \citep[for example,][]{Zhou20}. The interested reader is referred to the survey of \cite{Kazemi20} and references therein.

Again, it is emphasised that the objective of this paper is \textit{not} to claim that the RDPG is superior to competing models, but to provide guidelines for practitioners using RDPGs in their application domains, offering insights on the performance of these models for link prediction purposes.

\section{Random dot product graphs and \\ adjacency spectral embedding} \label{grdpg_sec} 

In this section, the generalised random dot product graph \citep{RubinDelanchy17} and methods for estimation of the latent positions are formally introduced. Suppose $\mvec A\in\{0,1\}^{n\times n}$ is a symmetric adjacency matrix of an undirected graph with $n$ nodes. 

\begin{definition}[{Generalised random dot product graph -- GRDPG}] \label{grdpg}
Let $\dplus$ and $\dminus$ be non-negative integers such that $d=\dplus+\dminus$. Let $\mathbb X\subseteq\mathbb R^{d}$ such that $\forall\ \vec x,\vec x^\prime\in\mathbb X$, $0\leq \vec x^\top{\mvec I}(\dplus,\dminus)\vec x^\prime\leq 1$, where
\begin{equation}
\mvec I(p,q) = \mathrm{diag}(1,\ldots,1,-1,\ldots,-1)
\end{equation}
with $p$ ones and $q$ minus ones.
Let $\mathcal{F}$ be a probability measure on $\mathbb{X}$, $\mvec A\in\{0,1\}^{n\times n}$ be a symmetric matrix and $\mvec{X}=(\vec x_1,\dots,\vec x_n)^\top\in\mathbb{X}^n$. Then $(\mvec A,\mvec X)\sim\mathrm{GRDPG}_{\dplus,\dminus}(\mathcal{F})$ if $\vec x_1,\dots,\vec x_n\overset{iid}{\sim}\mathcal F$ and for $i<j$, 
$\mathbb P(A_{ij}=1)=\vec x_i^\top{\mvec I}(\dplus,\dminus)\vec x_j$ independently. 
\end{definition}

The adjacency spectral embedding (ASE) provides consistent estimates of the latent positions in GRDPGs \citep{RubinDelanchy17}, up to indefinite orthogonal transformations. 

\begin{definition}[Adjacency spectral embedding -- ASE] \label{adj_emb}
For $d\in\{1,\ldots,n\}$, consider the spectral decomposition
$\mvec A = \hat{\mathbf\Gamma}\hat{\mathbf\Lambda}\hat{\mathbf\Gamma}^\top +  \hat{\mathbf\Gamma}_\perp\hat{\mathbf\Lambda}_\perp\hat{\mathbf\Gamma}_\perp^\top,$
where $\hat{\mathbf\Lambda}$ is a $d\times d$ diagonal matrix containing the top $d$ eigenvalues in magnitude, in decreasing order, $\hat{\mathbf\Gamma}$ is a $n\times d$ matrix containing the corresponding orthonormal eigenvectors, and the matrices $\hat{\mathbf\Lambda}_\perp$ and $\hat{\mathbf\Gamma}_\perp$ contain the remaining $n-d$ eigenvalues and eigenvectors. The adjacency spectral embedding $\hat{\mvec X}=[\hat{\vec x}_{1},\dots,\hat{\vec x}_{n}]^\top$ of $\mvec A$ in $\mathbb R^d$ is 
\begin{equation}
\hat{\mvec X} = \hat{\mathbf\Gamma}\vert\hat{\mathbf\Lambda}\vert^{1/2}\in\mathbb R^{n\times d},
\end{equation}
where the operator $\vert\cdot\vert$ applied to a matrix returns the absolute value of its entries.
\end{definition}

If the graph is directed, and the adjacency matrix is not symmetric, it could be implicitly assumed that the generating model is $\mathbb P(A_{ij}=1)=\vec x_i^\top\vec y_j, \vec x_i,\vec y_j\in\mathbb X$, where each node is given two different latent positions, representing the behaviour of the node as source or destination of the link. In this case, the embeddings can be estimated using the singular value decomposition (SVD).

\begin{definition}[Adjacency embedding of the directed graph -- DASE] \label{dase}
Given a directed graph with adjacency matrix $\mvec A\in\{0,1\}^{n\times n}$, and a positive integer $d$, $1\leq d\leq n$, consider the singular value decomposition 
\begin{equation}
\mvec A = \begin{bmatrix} \hat{\mvec U} & \hat{\mvec U}_\perp\end{bmatrix} \begin{bmatrix} \hat{\mvec D} & \mvec 0 \\ \mvec 0 & \hat{\mvec D}_\perp \end{bmatrix} \begin{bmatrix} \hat{\mvec V}^\top \\ \hat{\mvec V}^\top_\perp \end{bmatrix} = \hat{\mvec U}\hat{\mvec D}\hat{\mvec V}^\top + \hat{\mvec U}_\perp\hat{\mvec D}_\perp\hat{\mvec V}^\top_\perp,
\end{equation}
where $\hat{\mvec D}\in\mathbb R_+^{d\times d}$ is diagonal matrix containing the top $d$ singular values in decreasing order, $\hat{\mvec U}\in\mathbb R^{n\times d}$ and $\hat{\mvec V}\in\mathbb R^{n\times d}$ contain the corresponding left and right singular vectors, and the matrices $\hat{\mvec D}_\perp$, $\hat{\mvec U}_\perp$, and $\hat{\mvec V}_\perp$ contain the remaining $n-d$ singular values and vectors. The $d$-dimensional directed adjacency embedding of $\mvec A$ in $\mathbb R^{d}$, is defined as the pair
\begin{equation}
  \hat{\mvec X}=\hat{\mvec U}\hat{\mvec D}^{1/2}, \quad \hat{\mvec Y}=\hat{\mvec V}\hat{\mvec D}^{1/2}.
\end{equation}
The DASE can be also naturally extended to bipartite graphs. 
\end{definition}

\section{Dynamic link prediction in random dot product graphs} \label{dynamic_pred}

Given a time series of network adjacency matrices $\mvec A_1,\mvec A_2,\dots,\mvec A_T$, the objective is to correctly predict $\mvec A_{T+1}$.
The most common approach in the literature \citep{Sharan08, Scheinerman10, Dunlavy11} is to analyse a collapsed version $\tilde{\mvec A}$ of the adjacency matrices:
\begin{equation}
\tilde{\mvec A} = \sum_{t=1}^T \psi_{T-t+1}\mvec A_t, \label{collapsed_matrix}
\end{equation}
where $\psi_1,\dots,\psi_T\in\mathbb R$ is a sequence of weights. \cite{Scheinerman10} propose to consider an average adjacency matrix, setting $\psi_t=1/T\ \forall\ t=1,\dots,T$, which corresponds to the maximum likelihood estimate of $\mathbb E(\mvec A_t)$ if $\mvec A_1,\dots,\mvec A_T$ are sampled independently from the same $\mathrm{Bernoulli}(\mvec X\mvec X^\top)$ distribution. 
The main limitation of such a model is that it is assumed that the graphs do not display any temporal evolution. 
Furthermore, if \eqref{collapsed_matrix} is used, it is assumed that all the possible edges of the adjacency matrix follow the same dynamics. 
Obtaining the ASE $\hat{\mvec X}=[\hat{\vec x}_1,\dots,\hat{\vec x}_n]$ of $\tilde{\mvec A}$ leads to an estimate the scores:
\begin{equation}
\mvec S = \hat{\mvec X}\hat{\mvec X}^\top. \label{cwa}
\end{equation}
For simplicity, the inner product \eqref{cwa} is not weighted by the matrix $\mvec I(d_+,d_-)$, implicitly assuming $d_+=d$ and $d_-=0$. 
The estimation approaches in \eqref{collapsed_matrix} and \eqref{cwa} will be used as baselines for comparison with alternative methods for temporal link prediction techniques using RDPGs, which will be proposed and discussed in the remainder of this section. 
The proposed methods will be classified in the following two categories:
\begin{itemize}
\item averages of inner products of embeddings (AIP),
\item inner products of an average of embeddings (IPA).
\end{itemize}

First, it is possible to consider the individual ASE for each adjacency matrix $\mvec A_t$, and calculate an AIP score:
\begin{equation}
\mvec S_{\aip} = \frac{1}{T}\sum_{t=1}^T \hat{\mvec X}_t\hat{\mvec X}_t^\top. \label{prediction1}
\end{equation}

A second option is to obtain an averaged embedding $\bar{\mvec X}$ from $\hat{\mvec X}_1,\hat{\mvec X}_2,\dots,\hat{\mvec X}_T$, and use this for predicting the link probabilities.
This procedure is slightly more complex than \eqref{prediction1}, since the embeddings are invariant to orthogonal transformations: given an orthogonal matrix $\mathbf\Omega_t\in\mathbb R^{d\times d}$, $\mathbb E(\mvec A_t)=\mvec X_t\mvec X_t^\top=(\mvec X_t\mathbf\Omega_t)(\mvec X_t\mathbf\Omega_t)^\top$. Therefore, the embeddings $\hat{\mvec X}_1,\dots,\hat{\mvec X}_T$ only provide estimates of the corresponding latent positions up to an orthogonal transformation, which could vary for different values of $t$. 
Consequently, the embeddings must be suitably \textit{aligned} or \textit{registered}, before a direct comparison can be carried out. Discussion of a technique to align the embeddings is deferred to Appendix~\ref{procrustes_section}.
Assuming that an averaged embedding $\bar{\mvec X}$ is obtained, the matrix of IPA scores for prediction of $\mvec A_{T+1}$ is:
\begin{equation}
\mvec S_{\ipa} = \bar{\mvec X}\bar{\mvec X}^\top. \label{prediction2}
\end{equation}

Similar scoring mechanisms can be derived for the techniques of multiple graph inference described in Section~\ref{sec:introduction}, such as the omnibus embedding \citep{Levin17}, based on the the omnibus matrix: 
\begin{equation}
\tilde{\mvec A} = \begin{bmatrix} \mvec A_1 & \displaystyle{\frac{\mvec A_1+\mvec A_2}{2}} & \cdots & \displaystyle{\frac{\mvec A_1+\mvec A_T}{2}} \\[.75em] \displaystyle{\frac{\mvec A_2+\mvec A_1}{2}} & \mvec A_2 & \cdots & \displaystyle{\frac{\mvec A_2+\mvec A_T}{2}} \\ \vdots & \vdots & \ddots & \vdots \\ \displaystyle{\frac{\mvec A_T+\mvec A_1}{2}} & \displaystyle{\frac{\mvec A_T+\mvec A_2}{2}} & \cdots & \mvec A_T \end{bmatrix}. \label{omni}
\end{equation}
The ASE $\hat{\mvec X}$ of $\tilde{\mvec A}$ gives $T$ latent positions for each node. The individual estimates $\hat{\mvec X}_t=[\hat{\vec x}_{1t},\dots,\hat{\vec x}_{nt}]$ of the latent positions for the $t$-th adjacency matrix are represented by the submatrix formed by the estimates between the $((t-1)n+1)$-th and $tn$-th row of $\hat{\mvec X}$.
Then, from the time series $\hat{\mvec X}_1,\hat{\mvec X}_2,\dots,\hat{\mvec X}_T$ of omnibus embeddings, a matrix of scores can be obtained using either AIP \eqref{prediction1} or IPA \eqref{prediction2}. In this case, the individual embeddings are directly comparable and an alignment step is not required. On the other hand, the omnibus embedding cannot easily be updated when new graphs $\mvec A_{T+1}, \mvec A_{T+2}, \dots$ become available, since $\tilde{\mvec A}$ and the embedding must be recomputed for each new snapshot. 
The idea of an omnibus embedding can be also easily extended to directed and bipartite graphs, constructing the matrix $\tilde{\mvec A}$ analogously and then calculating the DASE. 

Embeddings generated using the more parsimonious COSIE model \citep{Arroyo19} are also considered. In COSIE networks, the latent positions are assumed to be common across the $T$ snapshots of the graph, but the link probabilities are scaled by a time-varying matrix $\mvec R_t\in\mathbb R^{d\times d}$: 
$\mathbb E(\mvec A_t) = \mvec X\mvec R_t\mvec X^\top. \label{cosie_eq}$
The common latent positions $\mvec X$ and the time series of weighting matrices $\mvec R_1,\dots,\mvec R_T$ can be estimated via multiple adjacency spectral embedding \citep[MASE,][]{Arroyo19}.

\begin{definition}[{Multiple adjacency spectral embedding -- MASE}] \label{mase_emb}
Given a sequence of network adjacency matrices $\mvec A_1, \dots, \mvec A_T$, and an integer $d\in\{1,\ldots,n\}$, obtain the individual ASEs $\hat{\mvec X}_t = \hat{\mathbf\Gamma}_t\vert\hat{\mathbf\Lambda}_t\vert^{1/2}\in\mathbb R^{n\times d}$. Then, construct the $n\times Td$ matrix 
$\tilde{\mathbf\Gamma} = [\hat{\mathbf\Gamma}_1,\dots,\hat{\mathbf\Gamma}_T]\in\mathbb R^{n\times Td}$, 
and consider its singular value decomposition 
$\tilde{\mathbf\Gamma} = \hat{\mvec U}\hat{\mvec D}\hat{\mvec V}^\top + \hat{\mvec U}_\perp\hat{\mvec D}_\perp\hat{\mvec V}^\top_\perp$,
where $\hat{\mvec D}\in\mathbb R_+^{d\times d}$ is a diagonal matrix containing the top $d$ singular values in decreasing order, $\hat{\mvec U}\in\mathbb R^{n\times d}$ and $\hat{\mvec V}\in\mathbb R^{Td\times d}$ contain the corresponding left and right singular vectors, and the matrices $\hat{\mvec D}_\perp$, $\hat{\mvec U}_\perp$, and $\hat{\mvec V}_\perp$ contain the remaining singular values and vectors. The $d$-dimensional multiple adjacency embedding of $\mvec A_1,\dots,\mvec A_T$ in $\mathbb R^d$ is given by $\hat{\mvec X}=\hat{\mvec U}$, which provides an estimate of $\mvec X$, and the sequence $\hat{\mvec R}_1,\dots,\hat{\mvec R}_T$, where
\begin{equation} 
\hat{\mvec R}_t = \hat{\mvec X}^\top \mvec A_t\hat{\mvec X}.
\end{equation}
\end{definition}

For prediction, the matrix of AIP scores could be obtained from the time series of estimated link probabilities $\hat{\mvec X}\hat{\mvec R}_1\hat{\mvec X}^\top,\dots,\hat{\mvec X}\hat{\mvec R}_T\hat{\mvec X}^\top$:
\begin{equation}
\mvec S_{\aip} = \frac{1}{T}\sum_{t=1}^T \hat{\mvec X}\hat{\mvec R}_t\hat{\mvec X}^\top. \label{cosie2}
\end{equation}

Alternatively, an averaged $\bar{\mvec R}$ could be equivalently obtained from the time series of estimates $\hat{\mvec R}_1,\dots,\hat{\mvec R}_T$. Combining $\bar{\mvec R}$ with the estimate of the latent positions $\hat{\mvec X}$ yields the IPA scores: 
\begin{equation}
\mvec S_{\ipa} = \hat{\mvec X}\bar{\mvec R}\hat{\mvec X}^\top. \label{cosie1}
\end{equation} 
The COSIE model can also be extended to directed and bipartite graphs assuming $\mathbb E(\mvec A_t)=\mvec X\mvec R_t\mvec Y^\top$. This construction leads to estimates 
$\hat{\mvec R}_t=\hat{\mvec U}^\top\mvec A_t\hat{\mvec V}$, where $\hat{\mvec U}$ and $\hat{\mvec V}$ are estimates of $\mvec X$ and $\mvec Y$ obtained from MASE on the DASE embeddings $\hat{\mvec X}_1,\dots,\hat{\mvec X}_T$ and $\hat{\mvec Y}_1,\dots,\hat{\mvec Y}_T$, based on two matrices $\tilde{\mathbf\Gamma}$ constructed from the left and right singular vectors (\textit{cf.} Definition~\ref{mase_emb}). 

In summary, several link prediction schemes based on random dot product graph models have been proposed, corresponding to three different types of spectral embedding:
\begin{itemize}
\item scores based on individual embeddings, \textit{cf.} \eqref{prediction1} and \eqref{prediction2},
\item omnibus scores, \textit{cf.} \eqref{prediction1} and \eqref{prediction2}, based on the matrix representation in \eqref{omni},
\item COSIE scores, \textit{cf.} \eqref{cosie2} and \eqref{cosie1}. 
\end{itemize}
Two scores, denoted AIP and IPA, are calculated for each embedding type.
All methods will be compared to the popular collapsed adjacency matrix method in \eqref{cwa}.
 
The methods described in this section are all based on truncated eigendecompositions of some form of the adjacency matrix. The full eigendecomposition of a $n\times n$ dense matrix requires a cubic computational cost $\mathcal O(n^3)$, but only $d$ eigenvectors and eigenvalues, where $d$ is in general $\mathcal O(1)$, are required in the algorithms presented in this section. This reduces the computational effort to $\mathcal O(n^2)$ \citep{Yang19}. Also, graph adjacency matrices are binary and normally highly sparse. In this setting, efficient algorithms based on the power method calculate the required decomposition of a matrix $\mvec A$ in $\mathcal O\{\mathrm{nnz}(\mvec A)\mathcal P(1/\varepsilon)\}$ \citep{Ghashami16}, where $\mathrm{nnz}(\cdot)$ denotes the number of non-zero entries of the matrix, $\mathcal P(\cdot)$ is a polynomial function and $\varepsilon\in(0,1)$ is an approximation error parameter. This allows the methodologies in this section to be scalable to large networks with the support of modern computer libraries.
For example, calculating the individual embeddings $\hat{\mvec X}_1,\dots,\hat{\mvec X}_T$ only has complexity $\mathcal O\{\mathcal P(1/\varepsilon)\sum_{t=1}^T\mathrm{nnz}(\mvec A_t)\}$. COSIE adds a further SVD decomposition in the MASE algorithm, which requires further $\mathcal O(nd^2)$ operations. On the other hand, calculating the omnibus embedding for large graphs might quickly become cumbersome, especially if $T$ is large, since up to $\mathcal O(n^2T^2)$ operations are required. 

\section{Improving prediction using \\ time series models} \label{ts_extension}

The collapsed matrix used in \eqref{collapsed_matrix} assumes that the underlying dynamics of each link are the same across the entire graph. This assumption is particularly limiting in real world applications, where different behaviours might be associated with different nodes or links. Instead, edge specific matrix parameters $\mathbf\Psi_1,\dots,\mathbf\Psi_T\in\mathbb R^{n\times n}$ might be able to more reliably capture the behaviour of each edge. 
A modification of the collapsed matrix $\tilde{\mvec A}$ in \eqref{collapsed_matrix} is therefore proposed: 
\begin{equation}
\tilde{\mvec A} = \sum_{t=1}^T \left(\mathbf\Psi_{T-t+1}\odot\mvec A_t\right), \label{collapsed_matrix_full}
\end{equation}
where $\mathbf\Psi_1,\dots,\mathbf\Psi_T\in\mathbb R^{n\times n}$ is a sequence of weighting matrices, and $\odot$ is the Hadamard element-wise product. 
The matrix in \eqref{collapsed_matrix_full} is denoted \textit{predicted adjacency matrix}. 
Note that in \eqref{collapsed_matrix_full}, the weights can only be estimated for those entries such that $A_{ijt}=1$ for at least one $t\in\{1,\dots,T\}$, but the ASE of $\tilde{\mvec A}$ still allows to estimate non-zero link probabilities even for those edges such that $A_{ijt}=0\ \forall t$. 

The idea could be easily extended to all the other prediction settings proposed in Section~\ref{dynamic_pred}, replacing the average link probability or average embedding with an autoregressive combination. For example, from the sequence of standard embeddings $\hat{\mvec X}_1,\hat{\mvec X}_2,\dots,\hat{\mvec X}_T$, it could be possible to obtain the scores as:
\begin{equation}
\mvec S_\text{PIP} = \sum_{t=1}^T \left(\mathbf\Psi_{T-t+1}\odot\hat{\mvec X}_t\hat{\mvec X}_t^\top\right). \label{prediction10}
\end{equation} 

Alternatively, it could be possible to use a similar technique to obtain an estimate $\tilde{\mvec X}_{T+1}=\sum_{t=1}^T (\mathbf\Psi_{T-t+1}\odot\hat{\mvec X}_t)$ of the embedding $\mvec X_{T+1}$, where in this case $\mathbf\Psi_t\in\mathbb R^{n\times d}$. The scores are then obtained as: 
\begin{equation}
\mvec S_\text{IPP} = \tilde{\mvec X}_{T+1}\tilde{\mvec X}_{T+1}^\top. \label{predict_proc}
\end{equation} 
Similarly, for COSIE, the scores could be based on a linear combination of $\hat{\mvec R}_1,\dots,\hat{\mvec R}_T$ to estimate $\mvec R_{T+1}$. 

The equations \eqref{prediction10} and \eqref{predict_proc} define two different methodologies to extend multiple RDPG inference models to a dynamic setting. 
Following the same nomenclature used in Section~\ref{dynamic_pred}, the scores based on time series models have been respectively denoted as PIP for the predicted inner product \eqref{prediction10}, and IPP for the inner product of the prediction \eqref{predict_proc}.
The PIP scores have a particularly interesting property: the node-specific embeddings are combined with time series models at the edge levels, fusing information at two different network resolutions.

Note that, for COSIE scores, $\mvec S_\text{AIP}=\mvec S_\text{IPA}$ from \eqref{cosie2} and \eqref{cosie1}, but $\mvec S_\text{PIP}\neq\mvec S_\text{IPP}$. This is because for $\mvec S_\text{PIP}$, the scores are obtained directly from a prediction based on the estimated link probabilities, whereas for $\mvec S_\text{IPA}$, the prediction is based on an estimate of $\mvec R_{T+1}$ from $\hat{\mvec R}_1,\dots,\hat{\mvec R}_T$.

For estimation of the weighting matrices $\mathbf\Psi_1,\dots,\mathbf\Psi_T$, the time series of link probabilities or node embeddings will be modelled independently. 
Seasonal autoregressive integrated (SARI) processes represent a flexible modelling assumption. A univariate time series $Z_1,\dots,Z_T\in\mathbb R$, is a $\text{SARI}(p,b)(P,B)_s$ with period $s$ if the series is a causal autoregressive process defined by the equation
\begin{equation}
\phi(L)\Phi(L^s) (1-L)^b(1-L^s)^B Z_t = \varepsilon_t,\ \varepsilon_t\overset{iid}{\sim}\mathbb N(0,\sigma^2), \label{arima}
\end{equation}
where $\phi(v)=1-\phi_1v-\ldots-\phi_pv^p$, $\Phi(v)=1-\Phi_1v-\ldots-\Phi_Pv^P$, and $L$ is the {\it lag} operator $L^kZ_t = Z_{t-k}$
\citep{BrockwellDavis}. 
For example, consider a process $\text{SARI}(1,0)(0,1)_s$. The model equation \eqref{arima} becomes $\tilde Z_t=\phi_1\tilde Z_{t-1}+\varepsilon_t$ with $\tilde Z_t=Z_t-Z_{t-s}$.
The process is causal if and only if $\phi(v)\neq0$ and $\Phi(v)\neq0$ for $\abs{v}\leq1$. 
The parameters of the process $p,b,P,B$ and $s$ are all \textit{integer-valued}, and can be interpreted as follows:
$s$ is the seasonal periodicity; $b$ corresponds to the differencing required to make the process stationary in mean, variance, and autocorrelations, and $B$ refers to the seasonal differencing; $p$ is to the number of autoregressive terms appearing in the equation, and similarly $P$ refers to the number of seasonal autoregressive terms. 
More details about SARI models and their interpretation are given in \cite{BrockwellDavis}.

The value of $s$ usually depends on the application domain. For example, in computer networks with daily network snapshots, it is reasonable to assume $s=7$, which represents a periodicity of one week. The remaining parameters, $p, b, P$ and $B$, could be estimated using information criteria. 
For small values of $T$, the corrected Akaike information criterion (AICc) is preferred. 
The corresponding coefficients of the polynomials $\phi(v)$ and $\Phi(v)$, and the variance $\sigma^2$, can be estimated via maximum likelihood \citep{BrockwellDavis}. 
For a discussion on automatic selection of the parameters in SARI models, see \cite{Hyndman08}.

For prediction of future values $Z_{t+1}$ conditional on $Z_1,\dots,Z_t$, the general forecasting equation is obtained from \eqref{arima}, setting $\varepsilon_{t+1}$ to its expected value $\mathbb E(\varepsilon_{t+1})=0$, and obtaining an estimate $\hat Z_{t+1}$ solving from the known terms of the equation. Analogously, $k$-steps ahead forecasts for $Z_{t+k}$ can also be obtained. 

In this article, the univariate time series $Z_1,\dots,Z_T$ modelled using SARI are of four different types:
\begin{itemize}
\item Time series of estimated link probabilities obtained from \textit{any} embedding method: for example $\hat{\vec x}_{i1}^\top\hat{\vec x}_{j1},\dots,\hat{\vec x}_{iT}^\top\hat{\vec x}_{jT}$ for the standard ASE, representing the sequence of estimated scores for the edge $(i,j)$. This type of time series is used to obtain PIP scores, see \eqref{prediction10};
\item Time series of node embeddings on a given embedding dimension: for example $\hat x_{ir1},\dots,\hat x_{irT}$, obtained considering only the $i$-th node embedding on the $r$-th dimension from $\hat{\mvec X}_1,\dots,\hat{\mvec X}_T$. Such values are used for the IPP scores, see \eqref{predict_proc};
\item Time series of entries of the COSIE matrix, used to obtain IPP scores, see \eqref{predict_proc}. For example: $\hat R_{kh1},\dots,\hat R_{khT}$, corresponding to the $(k,h)$-th entry in $\hat{\mvec R}_1,\dots,\hat{\mvec R}_T$; 
\item Times series of binary entries of the network adjacency matrices: for example $A_{ij1},\dots,A_{ijT}$ for the edge $(i,j)$, used for \eqref{collapsed_matrix_full}.
\end{itemize} 
The coefficients for each entry of the weighting matrices $\mathbf\Psi_1,\dots,\mathbf\Psi_T$ are obtained by matching \eqref{collapsed_matrix_full}, \eqref{prediction10} and \eqref{predict_proc} with the model equation \eqref{arima}.

In this work, the binary time series $\mvec A_1,\mvec A_2,\dots,\mvec A_T$ is also modelled using indipendent SARI models for estimation of \eqref{collapsed_matrix_full}. 
Such a modelling approach might not be entirely technically suited for binary-valued time series, but this choice has relevant practical advantages: most programming languages have packages for automatic estimation of the parameters in SARI models, whereas the choice of initial values and estimation of the parameters in most generalised process for binary time series \citep{MacDonald97,Kauppi08,Benjamin03} is notoriously difficult, which is not desirable when the estimation task should be performed automatically and in parallel over a large set of time series. 

The time series modelling extensions presented in this section are computationally expensive, since up to $n^2$ or $nd$ time series model are fitted, each with complexity $\mathcal O(Tk)$, where $k$ is the number of models compared for the purpose of model selection using AICc. This results in a computational cost of $\mathcal O(n^2Tk)$ for PIP scores or $\mathcal O(ndTk)$ for IPP scores, difficult to manage for large networks. Therefore, with the exception of the IPP COSIE scores, the methodologies proposed in this section do not scale well to large networks, unlike the techniques proposed in Section~\ref{dynamic_pred}. 
In the case of the IPP COSIE score \eqref{cosie1}, the cost for prediction of future values of $\mvec R_t$ is only $\mathcal O(d^2Tk)$, independent of $n$. 

\begin{figure*}[!t]
\centering
\includegraphics[width=\textwidth]{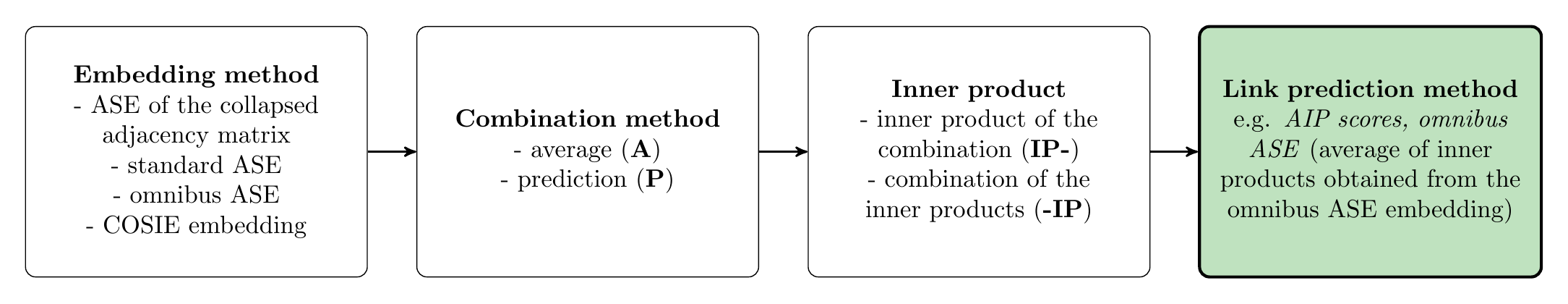}
\caption{Flowchart summarising the construction of a link prediction method using the techniques in Sections~\ref{dynamic_pred} and \ref{ts_extension}.}
\label{flow_scores}
\end{figure*}

The methodologies for constructing link prediction scores discussed in this work are summarised in Figure~\ref{flow_scores} in a flowchart. In summary, three choices must be made:
\begin{itemize}
\item embedding method (collapsed adjacency matrix, standard ASE, omnibus ASE, COSIE embedding);
\item combination method (average, \textit{cf.} Section~\ref{dynamic_pred}, or prediction, \textit{cf.} Section~\ref{ts_extension});
\item inner product (combination of inner products or inner product of a combination).
\end{itemize}

Clearly, the same methodology could be applied to \textit{any} embedding method, not necessarily based on the RDPG. Some examples will be given in Section~\ref{santander_alternative}.

\section{Results} \label{results}

The proposed methods were tested on synthetic data and on real world dynamic networks from different application domains: transportation systems, cyber-security, and co-authorship networks. For all examples, the parameter $d$ was selected using the profile likelihood elbow criterion of \cite{zhu}, unless otherwise specified.

\subsection{Simulated data} \label{sim_section}

\subsubsection{Seasonal stochastic blockmodel}

The performance of the rival link prediction techniques discussed in this article is initially compared on simulated data from stochastic blockmodels. The stochastic blockmodel \citep{Holland83} can be interpreted as a special case of a GRDPG \citep{RubinDelanchy17}: each node $i$ is assigned a latent community $z_i\in\{1,\dots,K\}$, with corresponding latent position $\bm\mu_{z_i}\in\mathbb R^d$; the probability of a link $(i,j)$ only depends on the community allocation of the two nodes:
$\mathbb P(A_{ij}=1) = \bm\mu_{z_i}^\top\mvec I(d_+,d_-)\bm\mu_{z_j}$.
To simulate a stochastic blockmodel, a within-community probability matrix $\mvec B=\{B_{ij}\}\in[0,1]^{K\times K}$ was generated, where $B_{ij}\sim\mathrm{Beta}(1.2,1.2)$ is the probability of a link between two nodes in communities $i$ and $j$, and $K$ is the number of communities. 
The matrix has full rank with probability 1, hence $K=d$. In the simulation, $T=100$ graph snapshots with $n=100$ and $K=5$ were generated. The community allocations were chosen to be time dependent, assuming a seasonality of one week. For each node, community allocations $z_{i,u},\ u=0,\dots,s-1$, with $s=7$, were sampled uniformly from $\{1,\dots,K\}$. Then, the adjacency matrices were obtained as:
\begin{equation}
\mathbb P(A_{ijt}=1) = B_{z_{i,t\bmod s},z_{j,t\bmod s}},\ t=1,\dots,T.
\end{equation}
Therefore, the link probabilities change over time, with a periodicity of $7$ days. The models presented in Section~\ref{dynamic_pred} were fitted using the first $T^\prime=80$ snapshots of the graph, with the objective of predicting the remaining $T-T^\prime$ adjacency matrices. The methods that are initially compared are:
\begin{itemize}
\item ASE of the \textit{averaged adjacency matrix}, \textit{cf.} \eqref{collapsed_matrix} and \eqref{cwa}, 
\item \textit{AIP and IPA scores} calculated from the \textit{standard ASE}, \textit{cf.} \eqref{prediction1} and \eqref{prediction2},
\item \textit{AIP and IPA scores} calculated from the \textit{omnibus embedding}, \textit{cf.} \eqref{omni},
\item \textit{AIP and IPA scores} calculated from \textit{COSIE}, \textit{cf.} \eqref{cosie2} and \eqref{cosie1}.
\end{itemize}

\begin{figure}[!t]
\centering
\includegraphics[width=.475\textwidth]{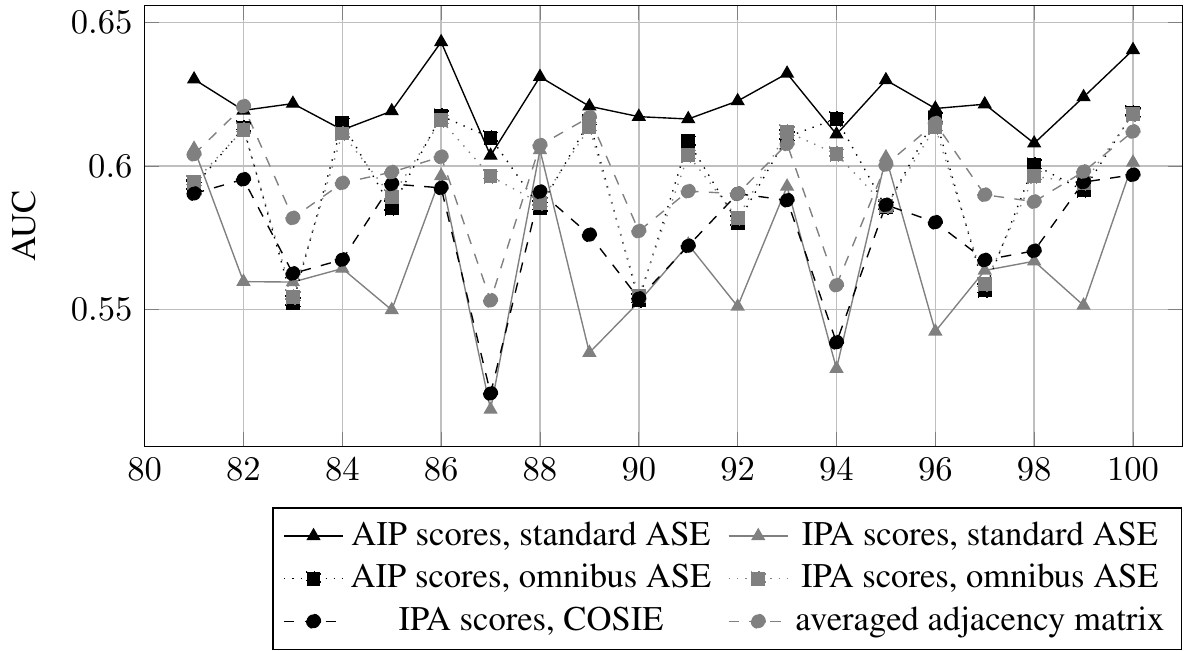}
\caption{Results of the link prediction procedure on the simulated seasonal SBM.}
\label{simulated_results}
\end{figure}

The link prediction problem can be framed as a binary classification task. Hence, the performances of the methods presented in this article are evaluated using the area under the receiver operating characteristic (ROC) curve, commonly known as AUC scores. The results are plotted in Figure~\ref{simulated_results}. 
The best performance is achieved by the AIP score based on the standard ASE, which outperforms the commonly used method of the collapsed adjacency matrix \eqref{collapsed_matrix}.

\begin{figure}[!t]
\centering
\begin{subfigure}[t]{0.475\textwidth}
\centering
\caption{Collapsed adjacency matrix scores \\ AIP and PIP scores, standard ASE}
\includegraphics[width=\textwidth]{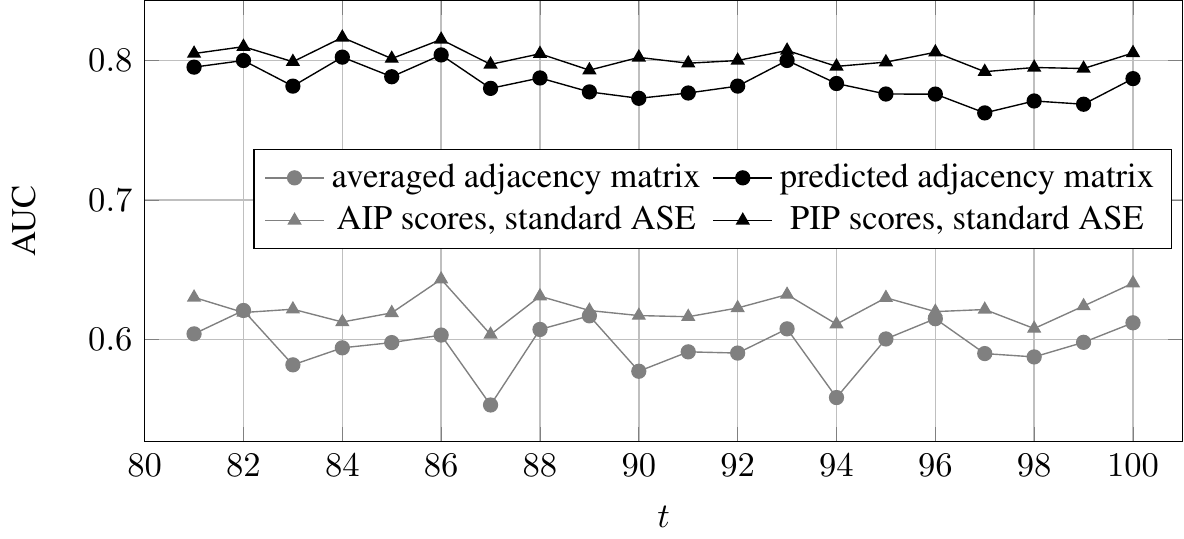}
\label{sim_collapsed}
\end{subfigure}
\begin{subfigure}[t]{0.475\textwidth}
\centering
\caption{IPA and IPP scores, standard ASE and COSIE}
\includegraphics[width=\textwidth]{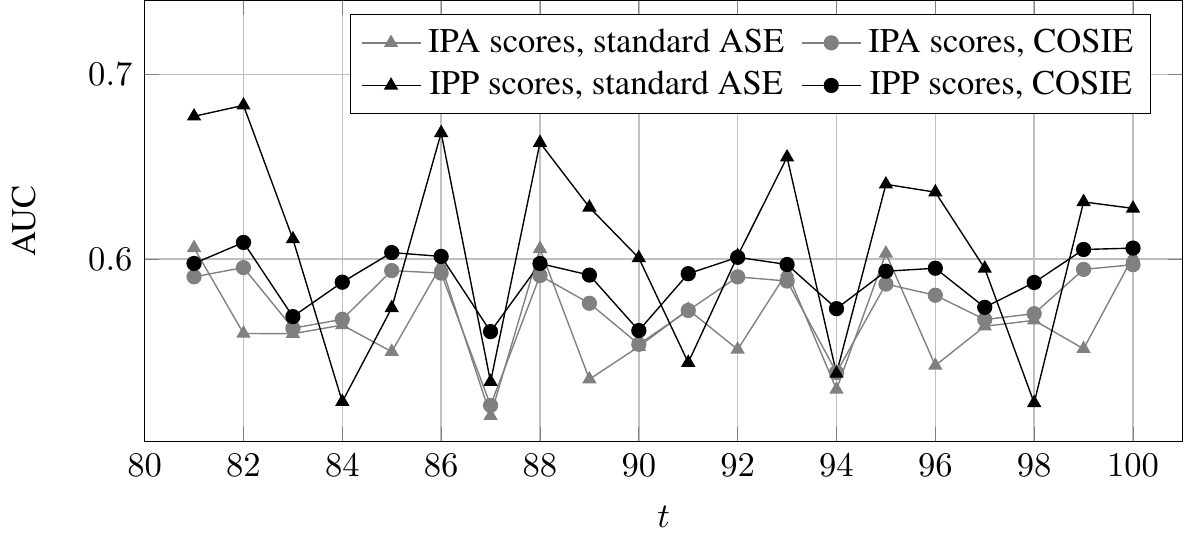}
\label{sim_aligned}
\end{subfigure}
\caption{Comparison between four of the link prediction models in Figure~\ref{simulated_results}, and their extensions using the methods in Section~\ref{ts_extension}, on the synthetic SBM data.}
\label{simulated_predicted}
\end{figure}

It is anticipated that the predictions should be improved upon by the PIP and IPP extensions presented in Section~\ref{ts_extension}, since the simulated network has clear dynamics which are not explicitly taken into account using the techniques from Section~\ref{dynamic_pred}. In particular, four methods are discussed:
\begin{itemize}
\item ASE of the \textit{predicted adjacency matrix}, \textit{cf.} \eqref{collapsed_matrix_full}, with weights obtained from independent seasonal $\mathrm{SARI}(p,b)(P,B)_7$ processes fitted on each binary sequence $A_{ij1},A_{ij2},\dots,A_{ijT^\prime}$ such that at least one $A_{ijt}=1$;
\item \textit{PIP scores} calculated from the \textit{standard ASE}, \textit{cf.} \eqref{prediction10}, based on the edge time series $\hat{\vec x}_{i1}^\top\hat{\vec x}_{j1},\hat{\vec x}_{i2}^\top\hat{\vec x}_{j2},\dots,\hat{\vec x}_{iT^\prime}^\top\hat{\vec x}_{jT^\prime}$ obtained from the individual ASEs on each $\mvec A_1,\mvec A_2,\dots,\mvec A_{T^\prime}$;
\item \textit{IPP scores} calculated from the \textit{standard ASE}, \textit{cf.} \eqref{predict_proc}, based on prediction of the subsequent embeddings $\tilde{\mvec X}_{T^\prime+1},\tilde{\mvec X}_{T^\prime+2},\dots$ from the time series of \textit{aligned}\footnote{The indefinite Procrustes alignment step, described in Appendix~\ref{procrustes_section}, has been implemented in \textit{python} using \textit{rpy2} and the R codebase developed by Joshua Agterberg, available online at \url{https://github.com/jagterberg/indefinite_procrustes} and \url{https://jagterberg.github.io/assets/procrustes_simulation.html}.} 
individual ASEs $\hat{\mvec X}_1,\dots,\hat{\mvec X}_{T^\prime}$;
\item \textit{IPP scores} calculated from \textit{COSIE}, based on prediction of correction matrices $\tilde{\mvec R}_{T^\prime+1},\tilde{\mvec R}_{T^\prime+2},\dots$ from the time series $\hat{\mvec R}_1,\dots,\hat{\mvec R}_{T^\prime}$, where independent models are fitted to the $d\times d$ time series corresponding to each entry.
\end{itemize}
The time series models were fit using the function \textit{auto\_arima} in the statistical \textit{python} library \textit{pmdarima}, using the corrected AIC criterion to estimate the number of parameters. The results are presented in Figure~\ref{simulated_predicted}. 

The AIP method \eqref{prediction1} which had the best performance in Figure~\ref{simulated_results}, is significantly improved by the PIP score \eqref{prediction10} using time series modelling, and it is overall the only method that reaches values of the AUC well above $0.8$.
Remarkably, the performance of the predicted adjacency matrix method in \eqref{collapsed_matrix_full} outperforms the results based on most of the other methods, despite the issues related to the modelling of binary time series pointed out in Section~\ref{ts_extension}. 
On the other hand, the improvements obtained using the COSIE-based scores and the IPP score \eqref{predict_proc} seem to be less significant compared to the two other methods. This aspect will also be confirmed on real data examples in the next section. 
In general, it is clear from the plots in Figure~\ref{simulated_predicted} that adding temporal dynamics to the network via time series modelling is beneficial for link prediction purposes. In particular, including edge specific information from the time series of estimated link probabilities, or from the binary time series of links, has significantly improved link prediction. 

\subsubsection{Logistic dynamic network model} \label{ldnm}

Next, the effect of the dynamic component on the prediction is evaluated. A directed dynamic graph with $n=100$ and $T=100$ is simulated, assuming  
$A_{ijt} \sim \mathrm{Bernoulli}(v_{ijt})$, $t=1,\dots,T$,
where 
\begin{equation}
\mathrm{logit}(v_{ijt}) = b_{ij} + c_{ij} \theta (t-1), \label{logimod}
\end{equation}
where $b_{ij}$ is a baseline, such that $b_{ij}\sim\mathrm{Uniform}(-6.9,0)$ independently for all pairs $(i,j)$, implying $v_{ij1}\in(0.001,0.5)$. Furthermore, $\theta\in\mathbb R$ is a trend parameter common to all the possible edges, and $c_{ij}\in\{-1,1\}$ is the sign of the trend on each edge, such that $\mathbb P(c_{ij}=1)=\mathbb P(c_{ij}=-1)=1/2$. 
Note that if $\theta=0$, the graph does not have any dynamics, whereas if $\abs{\theta}$ increases, the graph dynamics also increases. Note that, asymptotically for $t\to\infty$, $v_{ijt}\to0$ if $c_{ij}\theta\to-\infty$, and $v_{ijt}\to1$ if $c_{ij}\theta\to\infty$. Dynamic graphs are simulated for $\theta\in\{0,0.025,0.05,0.075\}$, and the \textit{AIP and PIP scores} obtained from \textit{standard ASE} trained on the first $T^\prime$ adjacency matrices are calculated for prediction of the last $T-T^\prime$ network snapshots. The results are then compared with the \textit{AIP scores} with \textit{standard ASE}, sequentially updated when new network snapshots become available, used for a 1-step ahead prediction of the next snapshot. 
If the network has a relevant dynamic component, the difference between the AUC obtained from the sequential and non-sequential AIP scores increases over time, because the network structure changes and the non-sequential scores cannot capture such evolution. On the other hand, the difference between the sequential AIP scores and the PIP scores should not show an increasing trend over time, since the time dynamics is taken into account via time series modelling.
 
\begin{figure}[!t]
\centering
\begin{subfigure}[t]{0.475\textwidth}
\centering
\caption{Difference between sequential and non-sequential AIP scores, standard ASE}
\includegraphics[width=\textwidth]{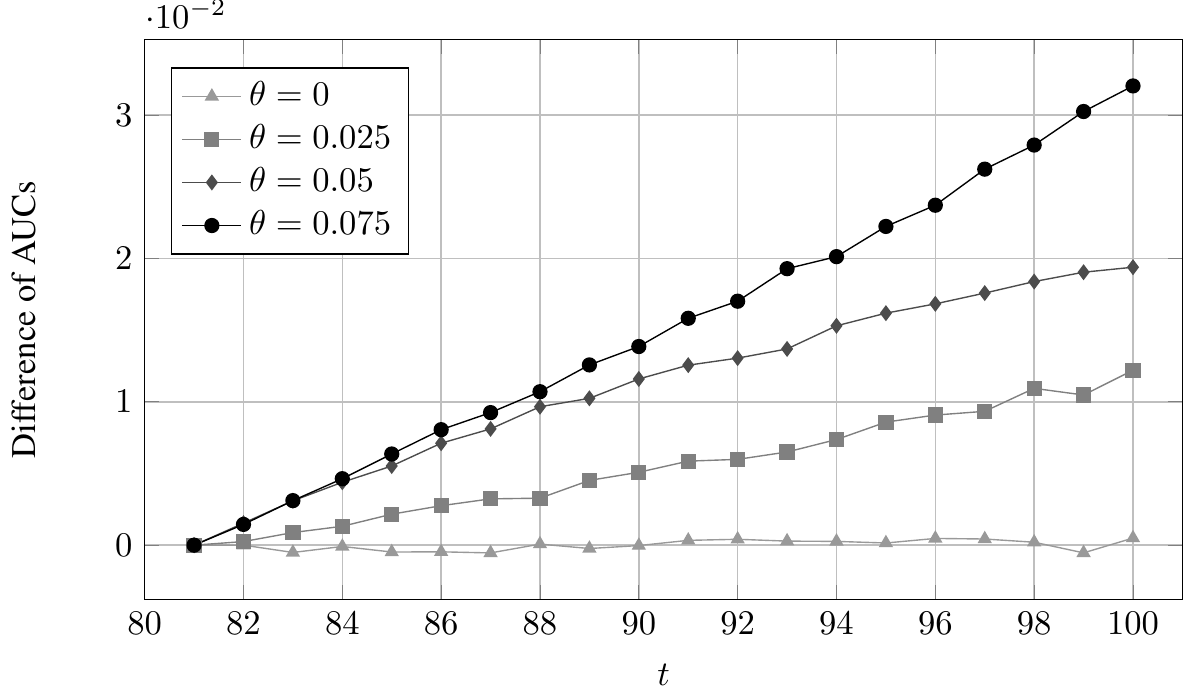}
\label{logistic_figure1}
\end{subfigure}
\begin{subfigure}[t]{0.475\textwidth}
\centering
\caption{Difference between sequential AIP and non-sequential PIP scores, standard ASE}
\includegraphics[width=\textwidth]{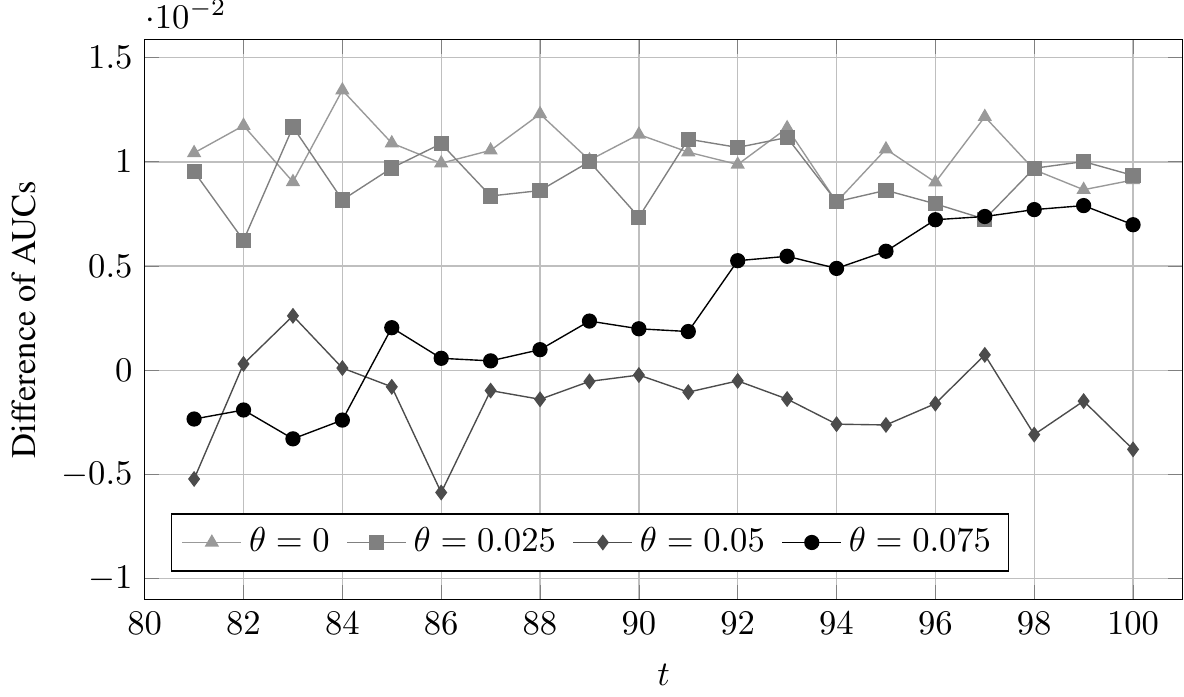}
\label{logistic_figure2}
\end{subfigure}
\caption{Difference between sequential AIP scores and (a) non-sequential AIP scores, (b) non-sequential PIP scores, calculated from standard ASE, for $\theta\in\{0,0.025,0.05,0.075\}$ in the logistic dynamic network model \eqref{logimod}.}
\label{logistic_figure}
\end{figure}

Figure~\ref{logistic_figure} plots the time series of differences between the sequential AIP scores and the corresponding non-sequential scores (Figure~\ref{logistic_figure1}), and the difference between the sequential AIP scores and the PIP scores (Figure~\ref{logistic_figure2}), for four different values of $\theta$. The plot demonstrates that in presence of a strong dynamic component, the sequential scores outperform non-sequential scores over time, as expected. On the other hand, the PIP scores perform similarly to the sequential scores, and the trend in the differences seem to disappear, up to fluctuations: this is overall remarkable, since the PIP scores are used for up to $(T-T^\prime)$-steps ahead predictions of 
matrices of scores based only on the initial $T^\prime$ adjacency matrices, whereas the sequential scores also use the last $T-T^\prime$ adjacency matrices sequentially for 1-step ahead predictions.

\subsection{Santander bikes} \label{santa}

Santander Cycles is a self-service cycle hire scheme operated in central London. Data about usage of the bikes are periodically released by Transport for London\footnote{The data are freely available at \url{https://cycling.data.tfl.gov.uk/}, powered by TfL Open Data.}. In this example, data from 7 March, 2018 to 19 March, 2019 were used, for a total of $T=378$ days. Each bike sharing station is considered as a node, and an undirected edge $(i,j,t)$ is drawn if at least one journey between the stations $i$ and $j$ is completed on day $t$. The total number of docking stations in London is $n= 840$. The daily graphs are fairly dense, with an average edge density of approximately $10\%$ across the $T$ networks. 
The first $T^\prime=250$ graphs are used as training set. 

\subsubsection{Averaged scores}

Initially, the methods compared are four of the techniques used to produce Figure~\ref{simulated_results}: 
\begin{itemize}
\item ASE of the \textit{averaged adjacency matrix}, 
\item \textit{AIP} and \textit{IPA} scores calculated from the \textit{standard ASE}, 
\item \textit{IPA} scores calculated from \textit{COSIE}. 
\end{itemize}

\begin{figure}[t]
\centering
\includegraphics[width=0.475\textwidth]{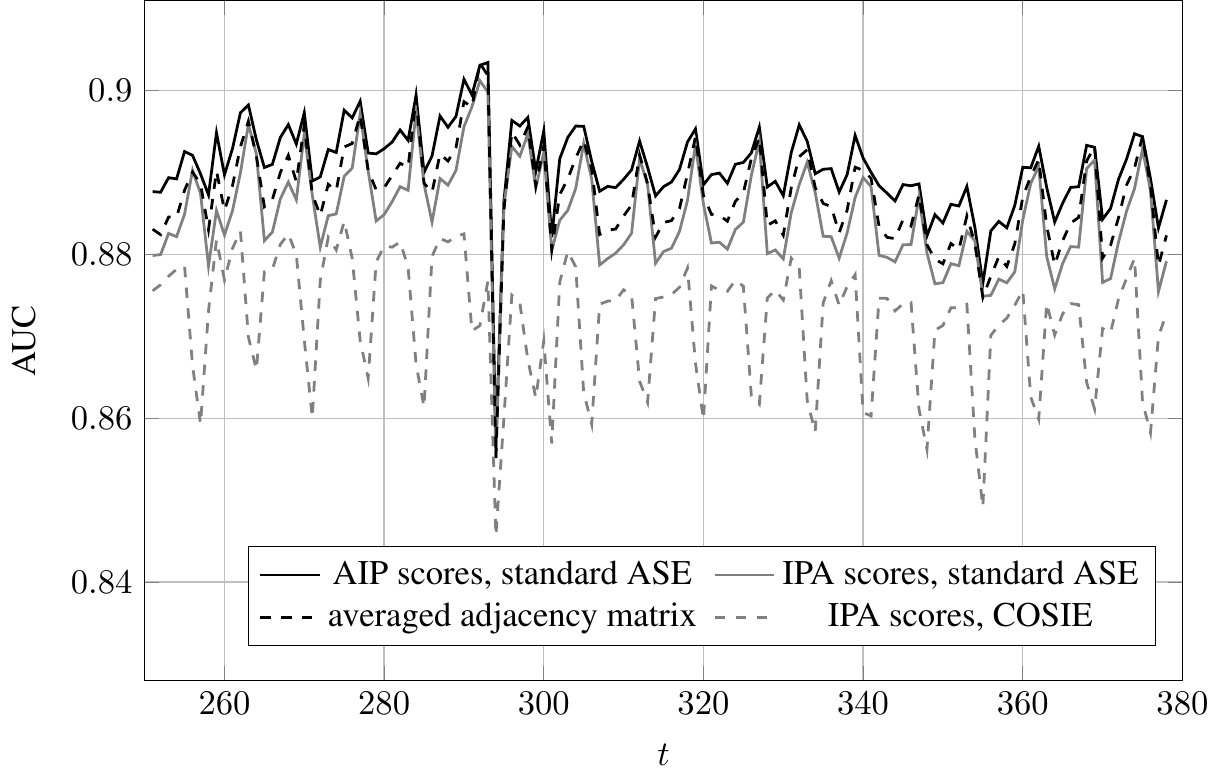}
\caption{Results of the link prediction procedure on the Santander Cycles network for different RDPG-based link prediction methods.}
\label{santander_results}
\end{figure}

For the Santander Cycles network, the results are reported in Figure~\ref{santander_results} for $d=10$. 
In Figure~\ref{santander_results}, the performance of the classification procedure drops around day $294$. This corresponds to Christmas day, which has a different behaviour compared to non-festive days. It is also interesting to point out that COSIE methods tends to perform better on weekdays than weekends, whereas the other methods more accurately predict the links on weekends compared to weekdays. 
The results of the link prediction procedure in Figure~\ref{santander_results} suggest that the data might not have a long term trend, but only a seasonal component, since the performance does not significantly decrease over time, and the parameters obtained using a training set of size $T^\prime=250$ seem to reliably predict the structure of the adjacency matrix even at $T=378$. 
Overall, this example seems to confirm that the method of AIP scores \eqref{prediction1} based on standard ASE
has the best performance for link prediction purposes when time dynamics are not included. 

\subsubsection{Comparison with alternative methods} \label{santander_alternative}

\begin{figure}[t]
\centering
\includegraphics[width=0.475\textwidth]{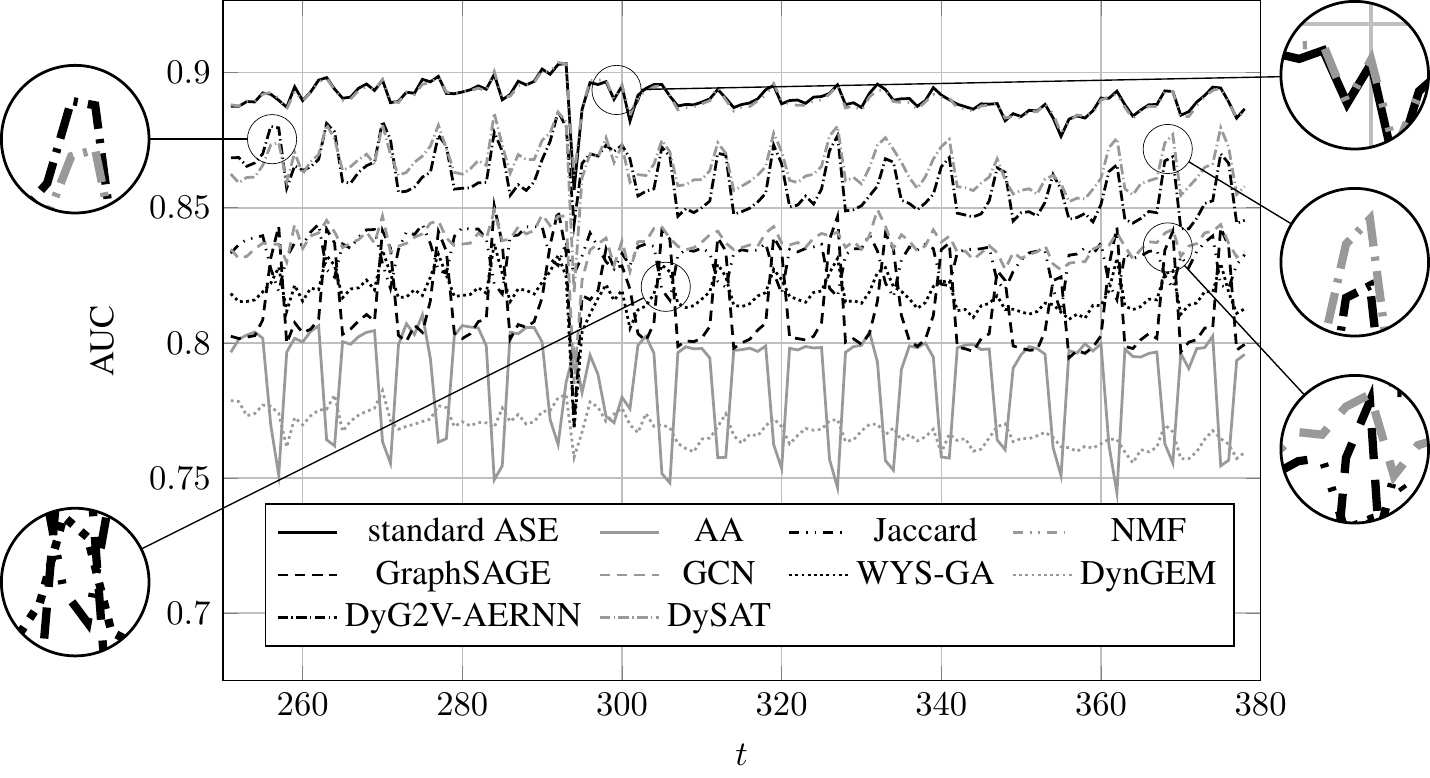}
\caption{Results of the link prediction procedure on the Santander Cycles network for different link prediction methods.}
\label{santander_results_other}
\end{figure}

To provide further comparison, the methods proposed in Section~\ref{dynamic_pred} are also compared in Figure~\ref{santander_results_other} to other methods used for link prediction in the literature. 
In order to demonstrate that the proposed methodology could be readily extended to any embedding technique, the embeddings were calculated from each of the adjacency matrices $\mvec A_1,\dots,\mvec A_T$, and prediction scores were obtained using the AIP methodology, akin to \eqref{prediction1}. 
The alternative methods considered in this section are:
\begin{itemize}
\item \textit{AIP scores} calculated from the \textit{Adamic-Adar} (AA) and \textit{Jaccard} coefficients \citep[used, for example, in][]{Gunes16}, 
\item \textit{AIP scores} calculated from \textit{non-negative matrix factorisation} \citep[NMF; see, for example,][]{Chen17,Yu17} with $d=10$, 
\item \textit{AIP scores} calculated from \textit{unsupervised GraphSAGE} \citep{Hamilton17}, \textit{GCN} \citep{Kipf17} with \textit{Deep Graph Infomax} \citep[DGI,][]{Veli19}, and \textit{Watch Your Step} with \textit{Graph Attention} \citep[WYS-GA,][]{Haija18}\footnote{The models were fitted using the implementation in the \textit{python} library \textit{stellargraph} \citep{StellarGraph}.}. Unsupervised network embeddings $\hat{\vec x}_{it}\in\mathbb R^d$ are obtained independently for each of the $T^\prime$ graphs in the training set, using one of the aforementioned methods with one-hot indicator vectors as node features (when required). 
For the unsupervised GraphSAGE, a two-layer model with sizes $50$ and $d=10$ is fitted, with 10 walks of length 10 per node, batch size 512 and 10 iterations in each encoder; Adam \citep{Kingma15} is used for learning the embeddings, with learning rate $10^{-2}$.  Note that the embedding dimension has been chosen to match the dimension of the RDGP-based embeddings. For the GCN, a one-layer network is trained, with layer size $d=10$ and ReLU activation, optimised by Adam with learning rate $10^{-2}$. 
For WYS-GS, $100$ walks per node are used, with $\beta=0.1$ and $C=10$ \citep[for the definition of such parameters, see][]{Haija18}, with embedding dimension $d=10$; the model is then trained with Adam with learning rate $10^{-3}$. 
For each of the methods, edge features are obtained from the Hadamard product $\hat{\vec x}_{it}\odot\hat{\vec x}_{jt}$ between the estimated node embeddings \citep[see, for example,][]{Grover16}. The link probabilities for each time window are then estimated from $T^\prime$ independent logistic regression models with response $A_{ijt}$ and $d$ predictors of the form $\hat{\vec x}_{it}\odot\hat{\vec x}_{jt}$. The link probabilities are then combined using the AIP method \eqref{prediction1}, and used to predict connections in the last $T-T^\prime$ observed graphs. 
\item Predictive scores calculated from three methods specifically developed for representation learning of dynamic graphs: the \textit{Deep Embedding Auto-Encoder Method for Dynamic Graphs} \citep[DynGEM,][]{Goyal18}, the \textit{dynamic graph2vec autoencoder recurrent neural network} model \citep[DyG2V-AERNN,][]{Goyal20}\footnote{The models were fitted using the implementation in the \textit{python} library \textit{DynamicGEM} \citep{Goyal_Code}.}, and the \textit{Dynamic Self-Attention Network} \citep[DySAT,][]{Sankar20}\footnote{The model was fitted using the code in the \textit{GitHub} repository \url{https://github.com/aravindsankar28/DySAT}.}. The methods were run using the default implementation of the software packages, setting the output embedding dimension to $d=10$ and the batch size to $100$. Link probabilities for the $T-T^\prime$ graphs in the test set were calculated using the same procedure described for GraphSAGE, GCN-DGI and WYS-GA.
\end{itemize}

The best performance among the alternative methods is achieved by the NMF scores, which achieve an almost equivalent performance to the AIP scores obtained from standard ASE. Slightly inferior performance is achieved by DyG2V-AERNN and DySAT, despite consistently exceeding $0.85$ in AUC.
GCN, Jaccard, WYS-GA and GraphSAGE have a slightly worse performance, but still consistently achieve AUC scores exceeding $0.8$. All of the proposed methodologies perform better than the Adamic-Adar and DynGEM, which seem to be largely outperformed by spectral embedding methods. Note that representation learning methods shine in particular when applied to large graphs with rich node features \citep[for example,][]{Hamilton17}, which is clearly not the case for the Santander cycles network. Furthermore, hyperparameter tuning is necessary to obtain a good link prediction performance, whereas RDGP-based methods only require the embedding dimension as input, and no further tuning is required. 

As discussed in Section~\ref{sec:introduction}, it should be noted that the methodologies of AIP and IPA scores proposed in Section~\ref{dynamic_pred}, and the corresponding PIP and IPP extensions in Section~\ref{ts_extension}, could be applied to \textit{any} sequence of individual embeddings, not necessarily obtained using ASE, but also with other embedding methods for static networks, for example NMF, as demonstrated in this section. 
Since one of the main objectives of this paper is comparing different embedding methods based on RDPGs, the focus in subsequent examples will be \textit{only} on RDPG-based techniques.

\subsubsection{Sequential scores}

\begin{figure}[t]
\centering
\includegraphics[width=0.475\textwidth]{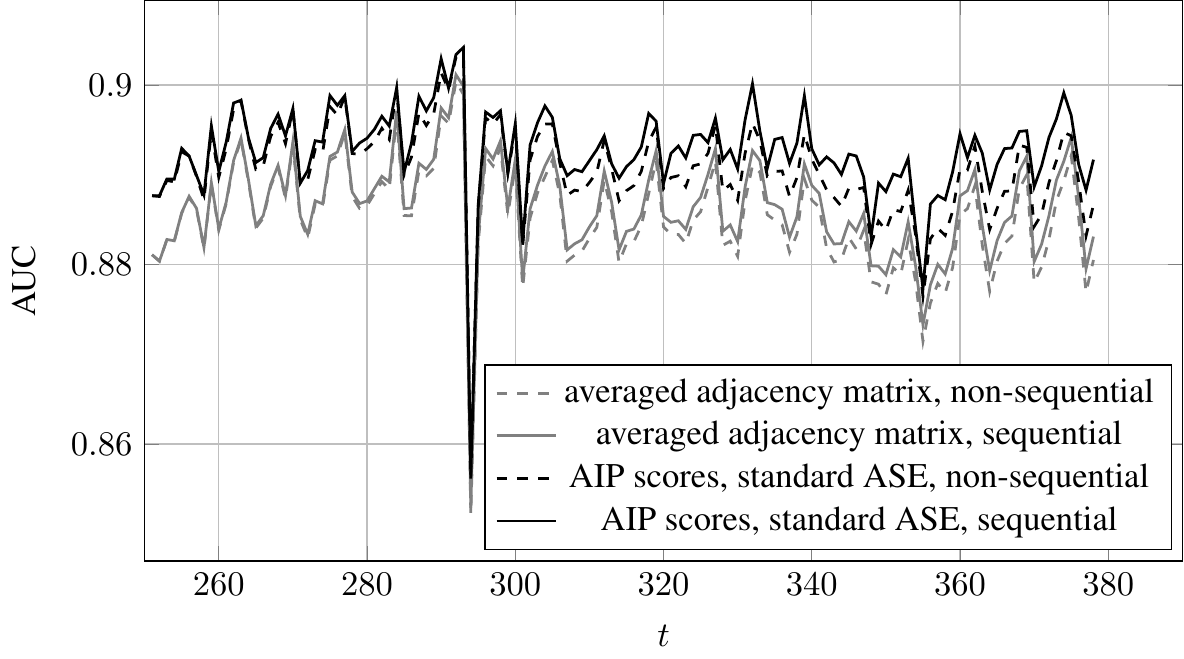}
\caption{Results for the AIP scores \eqref{prediction1} and averaged adjacency matrix scores for the Santander Cycles network, with and without sequential updates.}
\label{santander_sequential}
\end{figure}

So far, the embeddings learned using the initial $T^\prime$ snapshots have been used to predict \textit{all} the remaining $T-T^\prime$ adjacency matrices. In practical applications, it would be necessary to sequentially update the scores when new snapshots become available over time, improving the predictive performance. This is demonstrated in Figure~\ref{santander_sequential}, where the \textit{AIP scores} for \textit{standard ASE} and the \textit{average adjacency matrix scores} from Figure~\ref{santander_results} are compared with their sequential counterparts, obtained when the model is updated using each new observation $\mvec A_t$ in the test set. 
Clearly, updating the scores sequentially is beneficial, especially towards the end of the test set, whereas the difference between the methodologies is negligible in the initial snapshots of the test set. Both methods seem to reliably predict even $k$-steps ahead network snapshots, since the non-sequential curves in Figure~\ref{santander_sequential} are fairly close to their sequential counterparts.
The method of AIP scores based on standard ASE outperforms the scores based on the averaged adjacency matrix, commonly used in the literature, including sequential settings. 
Furthermore, the \textit{non-sequential} AIP scores \eqref{prediction1} also outperform the \textit{sequential} averaged adjacency matrix. This result is quite remarkable, since the former only use the initial $T^\prime$ snapshots of the network for training, whereas the latter is sequentially updated with the snapshots in the test set. 

\subsubsection{Predicted scores} \label{predicted_santander_section}

\begin{figure}[t]
\centering
\begin{subfigure}[t]{0.475\textwidth}
\centering
\caption{IPA and IPP scores, COSIE}
\includegraphics[width=\textwidth]{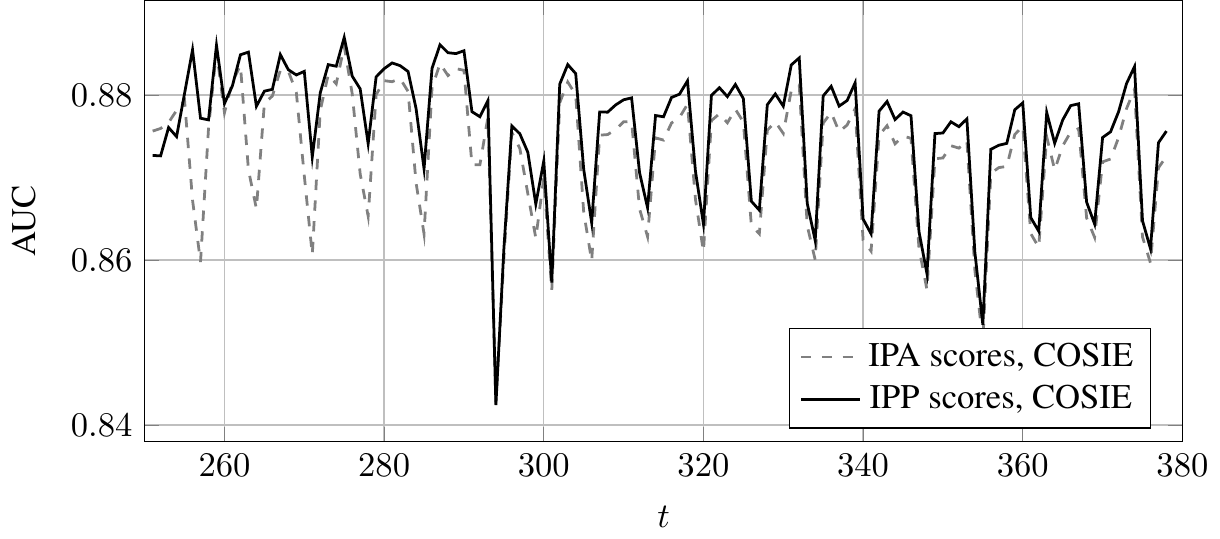}
\label{cosie_santander}
\end{subfigure}
\begin{subfigure}[t]{0.475\textwidth}
\centering
\caption{IPA and IPP scores, standard ASE}
\includegraphics[width=\textwidth]{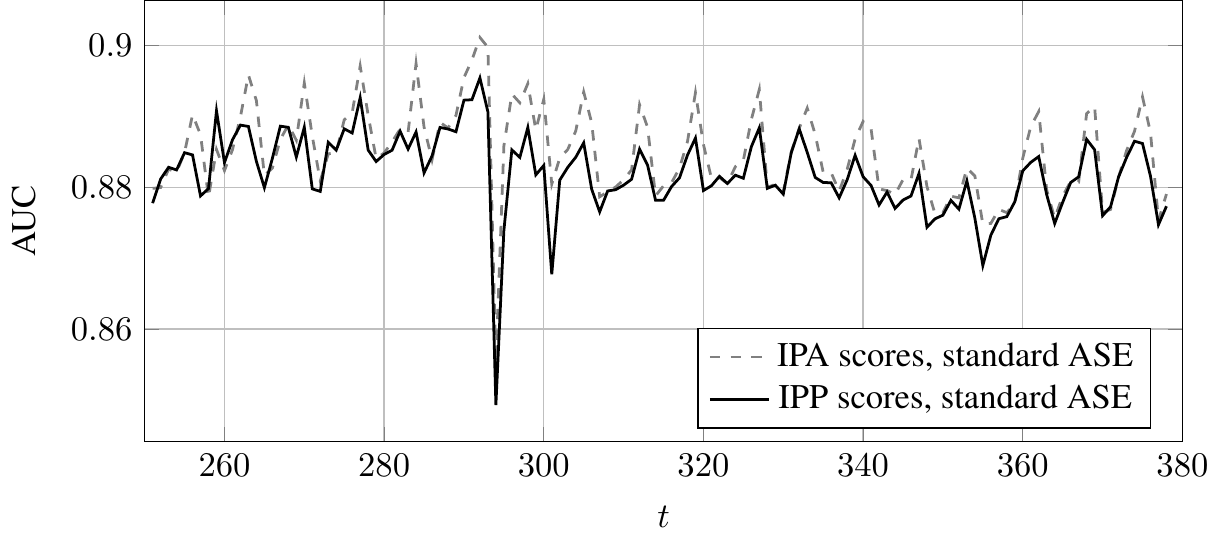}
\label{standard_santander}
\end{subfigure}
\caption{Comparison between two of the link prediction models in Figure~\ref{santander_results}, and their extensions using the methods in Section~\ref{ts_extension}, on the Santander Cycles network.}
\label{santander_predicted}
\end{figure}

The performance of the classifiers can be improved using some of the time series model in Section~\ref{ts_extension}. Figure~\ref{cosie_santander} show the results obtained from the prediction of subsequent COSIE correction matrices. The predictive performance is slightly improved by the extended time series models. Again, it is empirically confirmed that adding temporal dynamics is beneficial for the performance of random dot product graph based classifiers. 

On the other hand, predicting the subsequent adjacency spectral embeddings from the time series of aligned embeddings $\hat{\mvec X}_1,\hat{\mvec X}_2,\dots,\hat{\mvec X}_{T^\prime}$ does not seem to improve the predictive performance. The results are presented in Figure~\ref{standard_santander}, and confirms the findings in Figure~\ref{sim_aligned}, where the improvements on the simulated network were less significant compared to other methods. In this case, the time series models are not able to capture the dynamics of the aligned embeddings, and the predictive performance does not improve in AUC.

The limited improvements in the results seem to suggest that the network does not have a strong dynamic component. The tradeoff between performance and the computational effort required to fit multiple independent time series simultaneously, would suggest use of the AIP scores \eqref{prediction1} based on standard ASE in practical applications. 

\subsection{Los Alamos National Laboratory dataset} \label{lanl_section}

The unified host and network dataset \citep{Turcotte18} released by the Los Alamos National Laboratory (LANL) consists of a collection of network flow and host event logs generated from their machines running Microsoft Windows. From the host event logs, $90$ daily user-authentication bipartite graphs have been constructed, writing $A_{ijt}=1$ if the user $i$ initiates a connection authenticating to computer $j$, on day $t$. This graph is known as the \textit{user -- destination IP} graph. A total of $n_1=\numprint{12222}$ users, $n_2=\numprint{5047}$ hosts, and $\numprint{85020}$ pairs $(i,j)$ are observed, corresponding to approximately $0.137\%$ of all possible links. 

\subsubsection{Averaged scores and subsampling}

The first $T^\prime=56$ matrices are used as training set. Note that it is computationally difficult to calculate $n_1\times n_2$ scores for each adjacency matrix, and storing such large dense matrices in memory is also not efficient. Therefore, an estimate of the AUC can be obtained by subsampling the negative class at random from the zeroes in the test set adjacency matrices. Two subsampling techniques are used to construct the negative class for prediction of 
$\mvec A_t$:
\begin{enumerate}
\item[(1)] the negative class is constructed by sampling pairs $(i,j)$ such that $A_{ijt}=0$,
\item[(2)] the negative class contains randomly selected pairs $(i,j)$ such that $A_{ijt}=0$, \textit{and} all pairs $(i,j)$ such that $A_{ijt}=0$ and $A_{ijt^\prime}=1$ for at least $1$ value of $t^\prime\in\{1,\dots,T\}$.
\end{enumerate}

\begin{figure}[!t]
\centering
\begin{subfigure}[t]{0.475\textwidth}
\centering
\caption{Subsampling \textit{(1)}.}
\includegraphics[width=\textwidth]{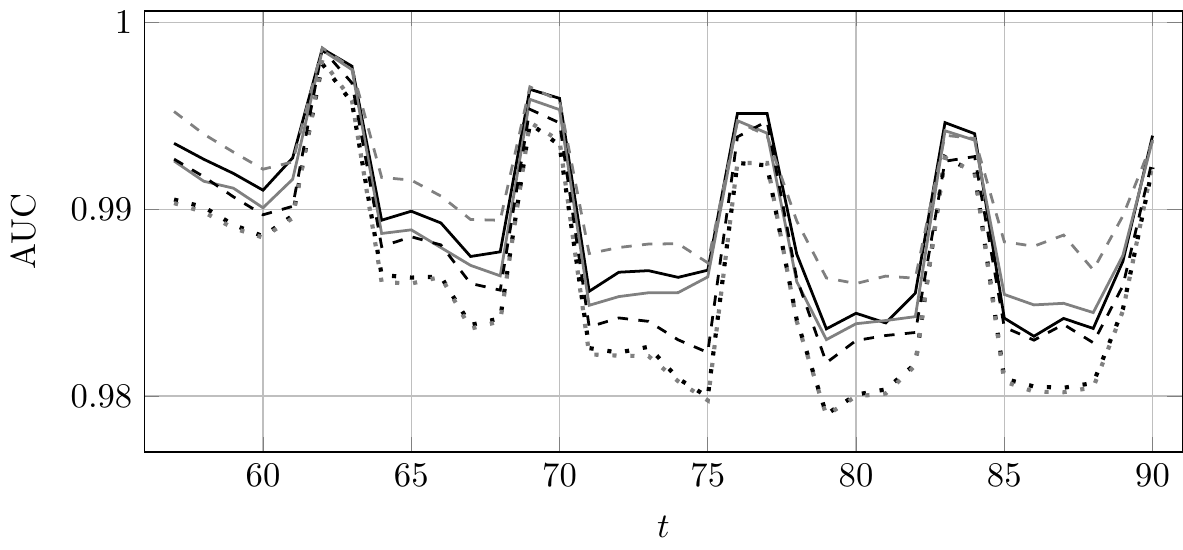}
\label{userdip_results1}
\end{subfigure}
\begin{subfigure}[t]{0.475\textwidth}
\centering
\caption{Subsampling \textit{(2)}.}
\includegraphics[width=\textwidth]{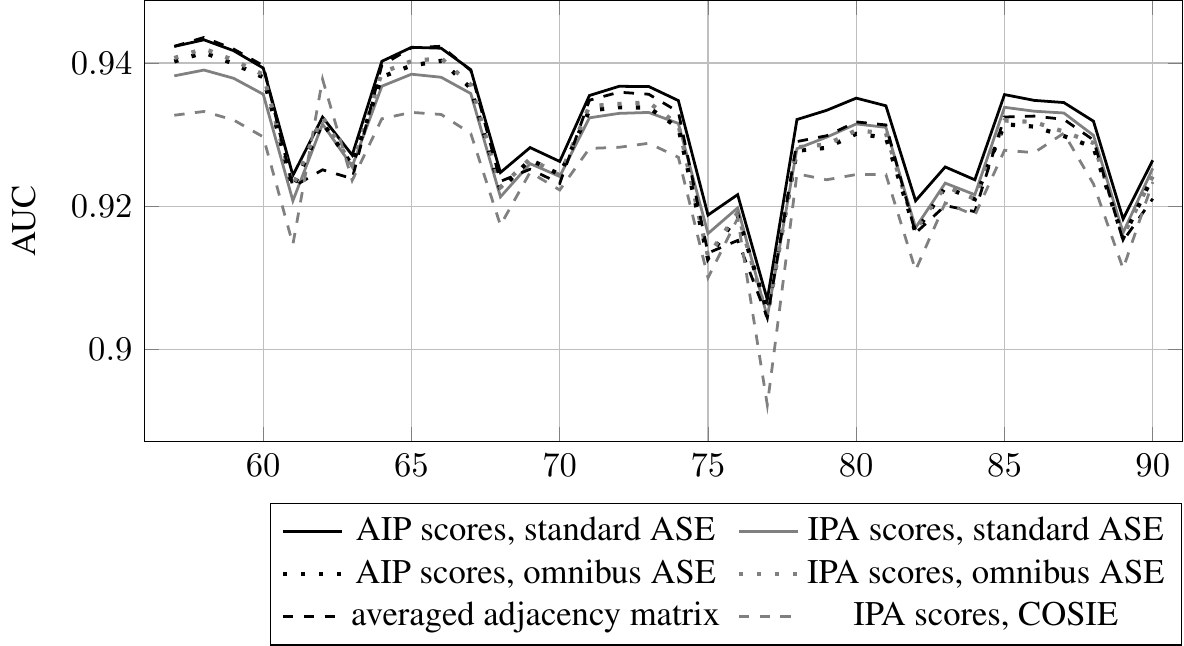}
\label{userdip_results2}
\end{subfigure}
\caption{Results of the link prediction procedure on the LANL network. AUCs are calculated from $\approx\numprint{150000}$ links per graph.}
\label{userdip_standard}
\end{figure}

For simplicity, the two techniques are denoted with the numbers \textit{(1)} and \textit{(2)} in Figure~\ref{userdip_standard}, which reports the results for the 6 methods considered in Figure~\ref{simulated_results} in Section~\ref{sim_section}.
The former subsampling technique provides an estimate of the ROC curve for the entire matrix, since the scores are sampled at random from the distribution of all scores. On the other hand, the latter method includes in the negative class more elements that tend to have associated high scores, represented by the pairs $(i,j,t)$ such that $A_{ijt}=0$ and $A_{ijt^\prime}=1$ for at least $1$ value of $t^\prime\in\{1,\dots,T\}$, giving an unbalanced sampling procedure and therefore a biased estimate of the ROC curve. 
Clearly, higher AUC scores are obtained using the first subsampling procedure.

Interestingly, in Figure~\ref{userdip_results1}, the COSIE-based AIP scores seem to have the best predictive performance across the different methods. In particular, COSIE scores tend to largely outperform the other methods during weekdays, whereas the performance during weekends seems almost equivalent, and sometimes inferior, to the AIP scores \eqref{prediction1} based on the standard ASE. 
On the other hand, in Figure~\ref{userdip_results2}, COSIE scores have the worst performance among the methods, except a spike on day $62$. In Figure~\ref{userdip_results2}, AIP scores \eqref{prediction1} based on the standard ASE
once again give the best predictive performance. The results of Figure~\ref{userdip_results2} are of particular interest since these allow for a comparison of the classifiers on a more challenging negative class compared to Figure~\ref{userdip_results1}. Therefore, it is reasonable to conclude that the AIP scores \eqref{prediction1} emerge again as the most suitable method for link prediction based on random dot product graphs.

\subsubsection{Significance of the difference between methods}

\begin{figure}[t]
\centering
\begin{subfigure}[t]{0.475\textwidth}
\centering
\caption{IPA scores, COSIE vs. \\ alternative methods -- Subsampling \textit{(1)}}
\includegraphics[width=\textwidth]{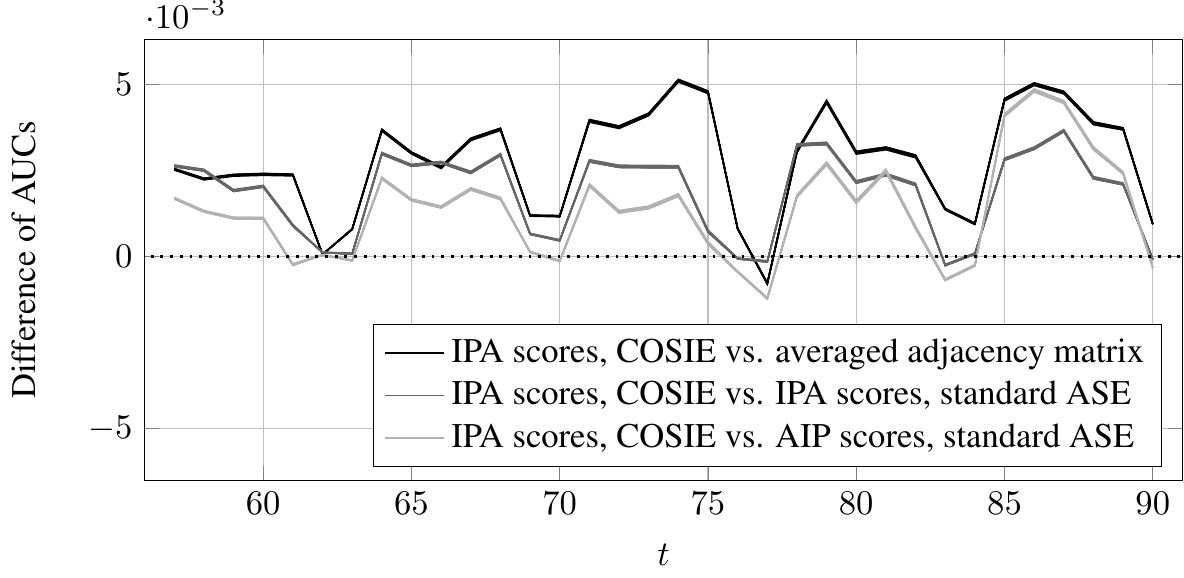}
\label{lanl_sign1}
\end{subfigure}
\begin{subfigure}[t]{0.475\textwidth}
\centering
\caption{AIP scores, standard ASE vs. \\ alternative methods -- Subsampling \textit{(2)}}
\includegraphics[width=\textwidth]{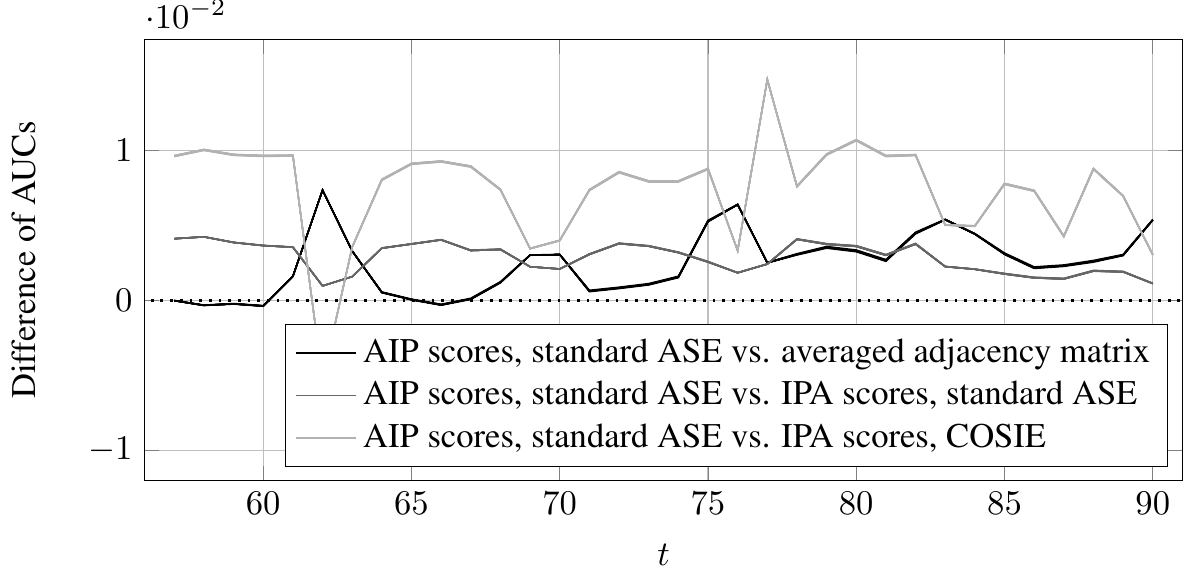}
\label{lanl_sign2}
\end{subfigure}
\caption{95\% confidence intervals of the difference between AUCs obtained alternative methods, obtained using subsampling \textit{(1)} in (a) and \textit{(2)} in (b). Reference method: (a) IPA scores, COSIE, (b) AIP scores, standard ASE.}
\label{lanl_sign}
\end{figure}

It could be argued that the differences between the methodologies do not appear to be significant, and might be due to the subsampling scheme. In order to assess significance of the differences, the subsampling procedure was repeated for $M=100$ times, and the corresponding AUCs were calculated. Figure~\ref{lanl_sign} plots the estimated 95\% confidence intervals for the difference between the AUCs obtained using different methods. Figure~\ref{lanl_sign1} uses the IPA scores calculated with COSIE as reference, and the AUCs are obtained with subsampling \textit{(1)}. Similarly, Figure~\ref{lanl_sign2} uses subsampling \textit{(2)}, and the AIP scores calculated with standard ASE are used as reference. Note that the width of the confidence intervals in Figure~\ref{lanl_sign} is barely visible, since the standard deviations of the AUC scores are $<10^{-4}$. This is because the number of subsamples is large enough to obtain extremely precise estimates of the AUC. Since the confidence intervals do not contain zero at any of the time points, the pairwise differences between the performance of different methodologies appear to be statistically significant. Similar confidence intervals, and corresponding $p$-values $<10^{-6}$ were observed for all the comparisons in the current and next sections, suggesting that the subsampling procedure for estimation of the AUC is robust.

\subsubsection{Streaming link prediction and sequential scores} \label{stream_scores}

For this example, it is also demonstrated how the proposed methods can be extended for streaming applications.
For example, the method of combining inner products of individual embeddings is particularly suitable for implementation in a streaming fashion, since the inner products are easily updated on the run when new snapshots of the graph are observed. A demonstration using the method of \textit{fixed forgetting factors} \citep{Gama13} is given here. The matrix of scores $\mvec S_t$ for prediction of $\mvec A_{t+1}$ is updated in streaming as follows:
\begin{align}
&w_t = \lambda w_{t-1} + 1,\\ 
&\mvec S_t = \left(1 - \frac{1}{w_t}\right)\mvec S_{t-1} + \frac{1}{w_t} \hat{\mvec X}_t\hat{\mvec X}_t^\top, \label{fff}
\end{align}
starting from $w_1=1$ and $\mvec S_1= \hat{\mvec X}_1\hat{\mvec X}_1^\top$. The forgetting factor $\lambda\in[0,1]$ is usually chosen to be close to $1$ \citep{Gama13}. Note that $\lambda=1$ corresponds to the sequential AIP scores in \eqref{prediction1}, calculated as in Figure~\ref{santander_sequential}, whereas smaller values of $\lambda$ give more weight to recent observations. In particular, $\lambda=0$ only gives weight to the most recent snapshot. 
Schemes similar to \eqref{fff} could be also implemented to update the collapsed adjacency matrix \eqref{collapsed_matrix}, obtaining the scores from its ASE at each point in time, or to update the matrix $\mvec R_t$ in \eqref{cosie_eq} for the COSIE embeddings, or the rotated embeddings $\hat{\mvec X}_t$.
The forgetting factor approach might be interpreted as a simplification of the time series scheme proposed in Section~\ref{ts_extension}, where the same weight is given to each edge. 

\begin{figure}[!t]
\centering
\begin{subfigure}[t]{0.475\textwidth}
\centering
\includegraphics[width=\textwidth]{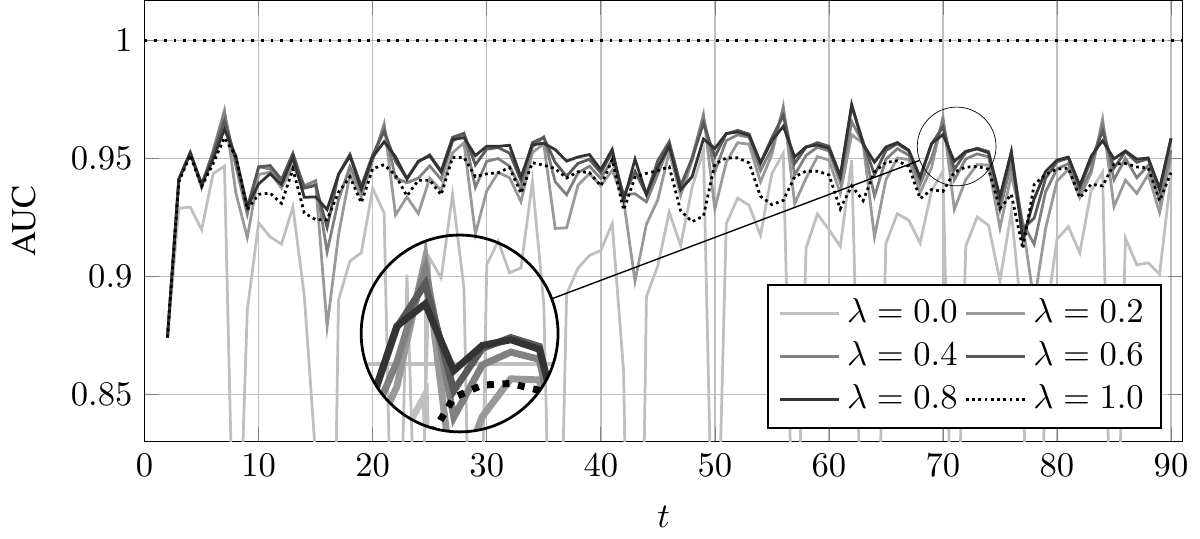}
\end{subfigure}
\caption{Results for the AIP scores \eqref{prediction1} on the LANL network with streaming updates. AUCs are calculated from $\approx\numprint{150000}$ links, using subsampling \textit{(2)}.}
\label{userdip_seqq}
\end{figure}

\begin{figure}[!t]
\centering
\includegraphics[width=0.475\textwidth]{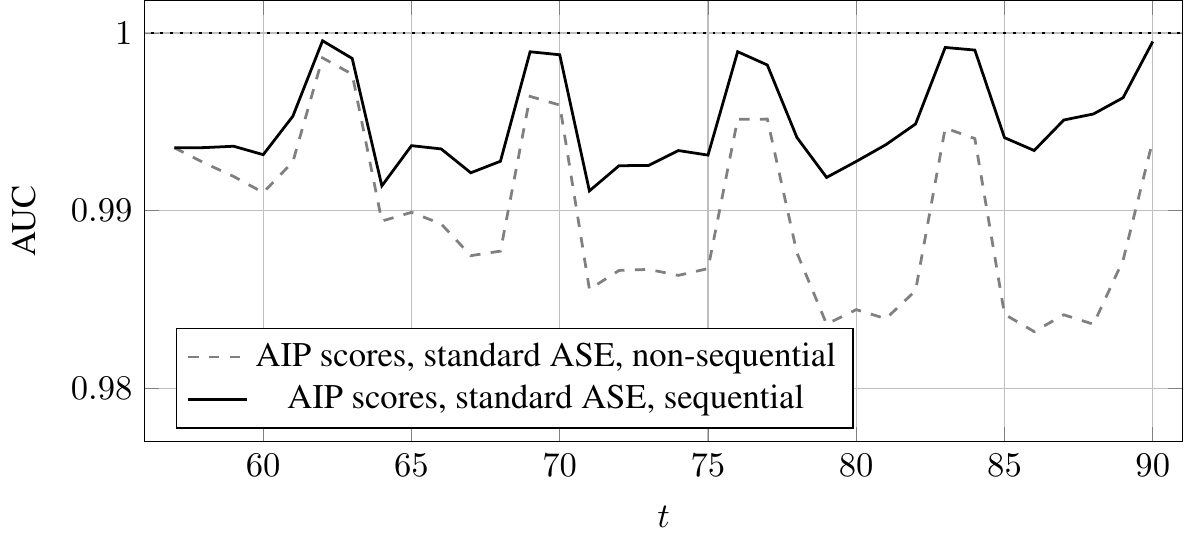}
\caption{Results for the AIP scores \eqref{prediction1} for the LANL network, with and without sequential updates. AUCs are calculated from $\approx\numprint{150000}$ links per graph, using subsampling \textit{(1)}.}
\label{userdip_sequential}
\end{figure}

The results for the entire observation period for a range of different values of $\lambda$ are plotted in Figure~\ref{userdip_seqq}. 
The best performance is achieved with the forgetting factor approach with $\lambda\in[0.4,0.8]$. The performance for $\lambda=0$ clearly drops around the weekends, since the network has a seasonal component which is not accounted for in the prediction. 
The difference between the curves for $\lambda=1$ and $\lambda<1$ suggests that the graph has temporal dynamics which is captured by the forgetting factor approach, which down-weights past observations in favour of more recent snapshots of the graph. 
This impression is confirmed by Figure~\ref{userdip_sequential}, which shows the sequential and non-sequential AIP scores based on the standard ASE, akin to Figure~\ref{santander_sequential}. 
The predictive performance slightly deteriorates for snapshots that are further away in time, whereas 1-step ahead predictions based on the sequential scores consistently give better results. 

\subsubsection{Predicted scores} \label{lanl_predicted_scores}

\begin{figure*}[!t]
\centering
\begin{subfigure}[t]{0.475\textwidth}
\centering
\caption{Collapsed adjacency matrix \\ Subsampling \textit{(1)}}
\includegraphics[width=0.95\textwidth]{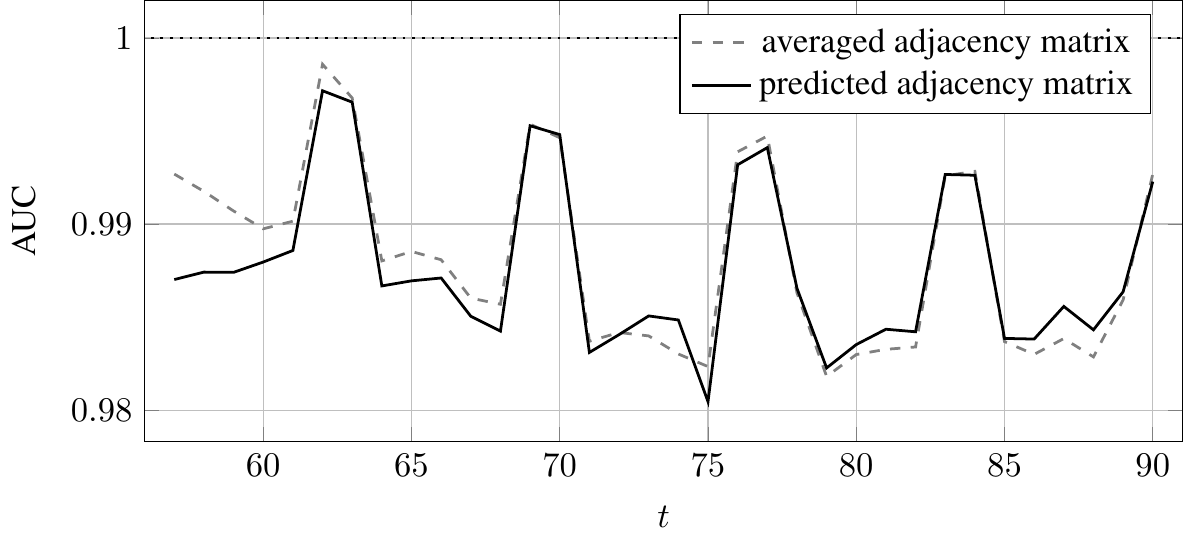}
\label{userdip_binary}
\end{subfigure}
\begin{subfigure}[t]{0.475\textwidth}
\centering
\caption{Collapsed adjacency matrix \\ Subsampling \textit{(2)}}
\includegraphics[width=0.95\textwidth]{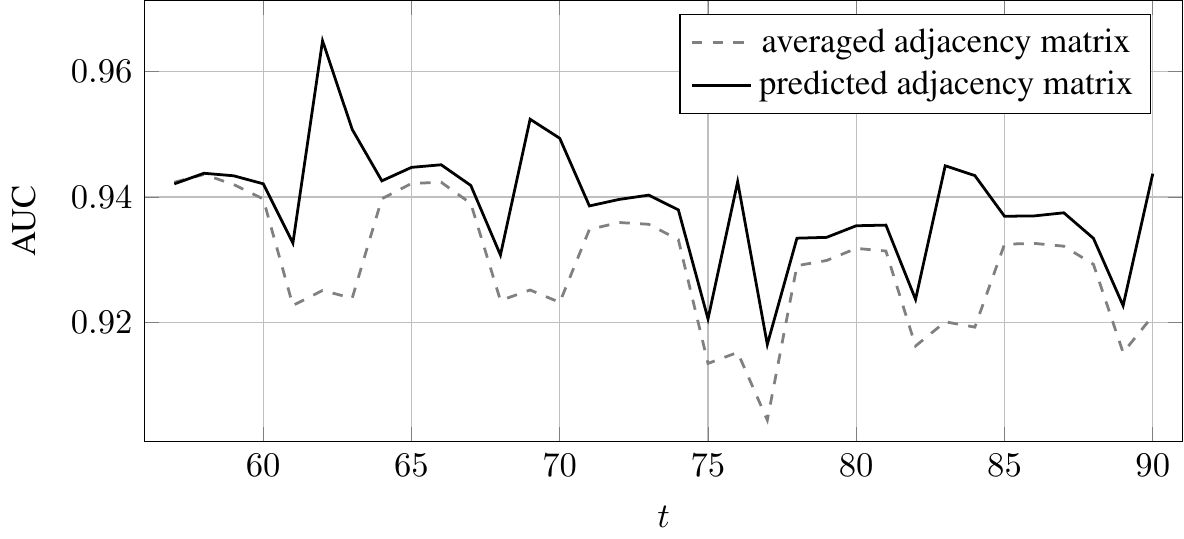}
\label{userdip_binary2}
\end{subfigure}
\begin{subfigure}[t]{0.475\textwidth}
\centering
\caption{AIP and PIP, standard ASE \\ Subsampling \textit{(1)}}
\includegraphics[width=0.95\textwidth]{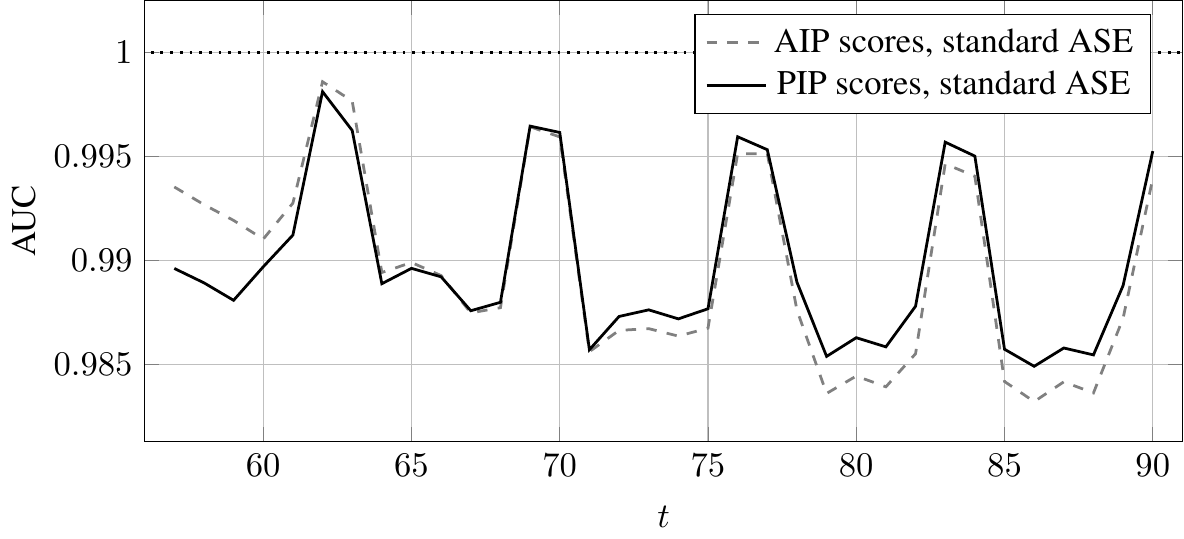}
\label{userdip_standardd}
\end{subfigure}
\begin{subfigure}[t]{0.475\textwidth}
\centering
\caption{AIP and PIP, standard ASE \\ Subsampling \textit{(2)}} 
\includegraphics[width=0.95\textwidth]{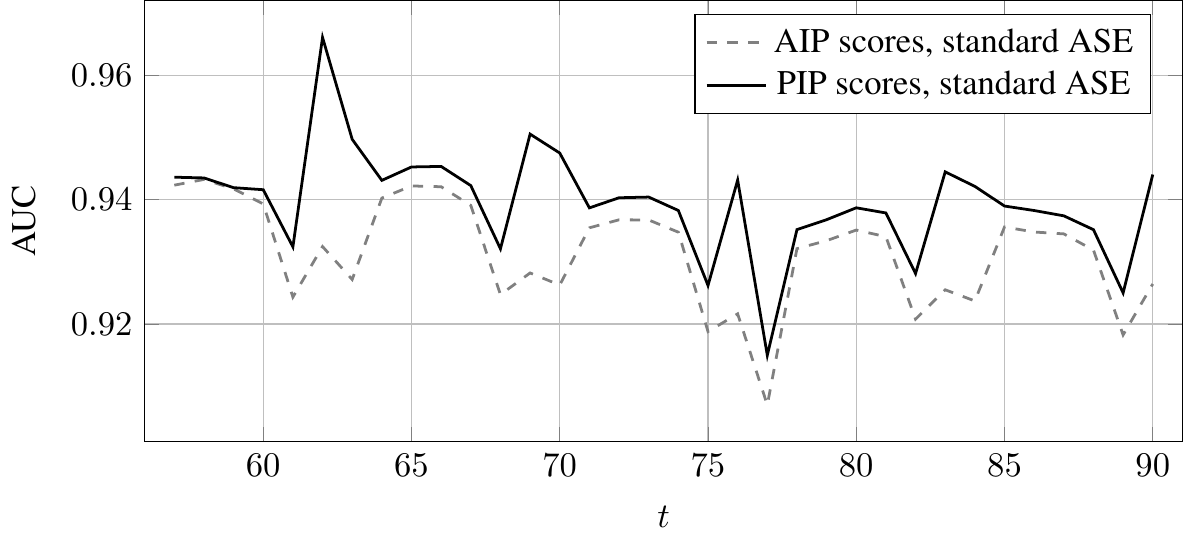}
\label{userdip_standard2}
\end{subfigure}
\begin{subfigure}[t]{0.475\textwidth}
\centering
\caption{IPA and IPP, COSIE scores \\ Subsampling \textit{(1)}}
\includegraphics[width=0.95\textwidth]{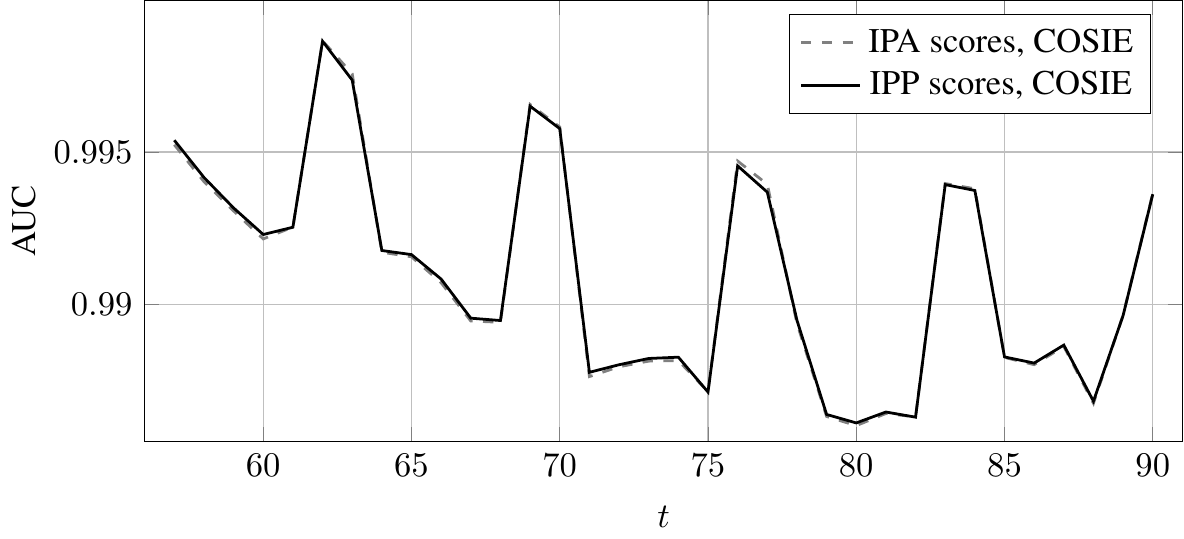}
\label{userdip_cosie}
\end{subfigure}
\begin{subfigure}[t]{0.475\textwidth}
\centering
\caption{IPA and IPP, COSIE scores \\ Subsampling \textit{(2)}}
\includegraphics[width=0.95\textwidth]{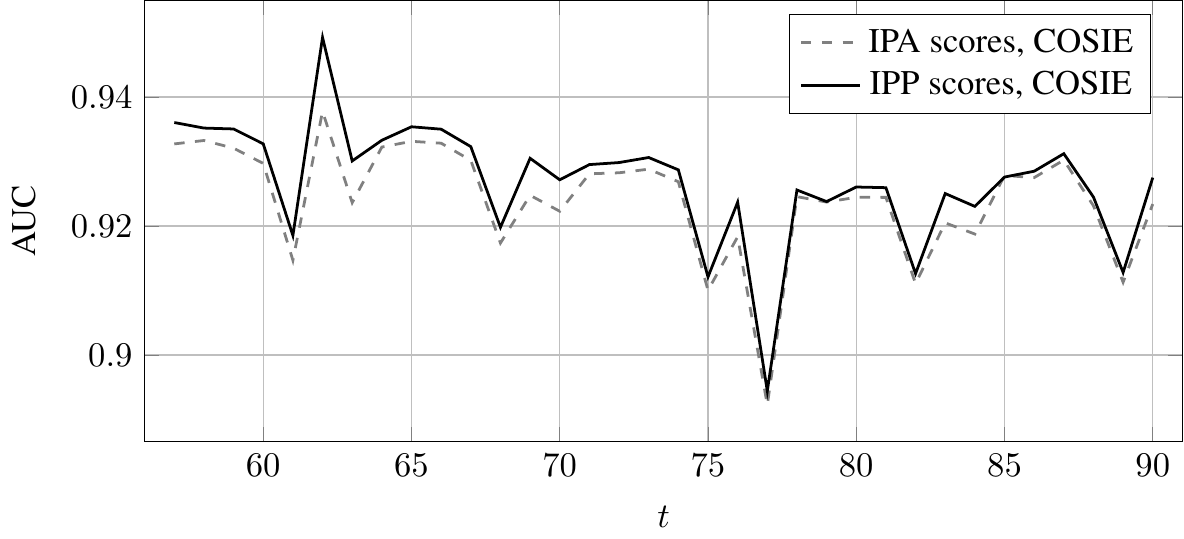}
\label{userdip_cosie2}
\end{subfigure}\caption{Comparison between three of the link prediction models in Figure~\ref{userdip_standard}, and their extensions using the methods in Section~\ref{ts_extension}, on the LANL network. AUCs are calculated from $\approx\numprint{150000}$ links per graph.}
\label{userdip_predicted}
\end{figure*}

The performance can be again improved using the time series models in Section~\ref{ts_extension}. Figure~\ref{userdip_predicted} shows the results obtained using the two different subsampling schemes for three methods:
\begin{itemize}
\item ASE of the \textit{averaged} and \textit{predicted adjacency matrix},
\item \textit{AIP} and \textit{PIP} scores calculated from \textit{standard ASE},
\item \textit{IPA} and \textit{IPP} scores calculated from \textit{COSIE}.
\end{itemize} 
From Figures~\ref{userdip_binary}, \ref{userdip_standardd} and \ref{userdip_cosie}, it is evident that the extensions do not improve the performance of the classifier when the subsampling scheme \textit{(1)} is used. On the other hand, Figures~\ref{userdip_binary2}, \ref{userdip_standard2} and \ref{userdip_cosie2} show that relevant improvements (especially on day $62$) are obtained when the subsampling method \textit{(2)} is used, which represents a more difficult classification task. Again, the results confirm that the performance of RDPGs for link prediction can be enhanced by time series models.

\subsection{Imperial College network flow data} \label{icl_section}

The methodologies proposed in Section~\ref{dynamic_pred} are also tested on a large computer network, demonstrating that the spectral embedding techniques are scalable to graphs of large size. A directed graph with $n=\numprint{95220}$ has been constructed using network flow data collected at Imperial College London, limiting the connections to internal hosts only. The data have been collected over $T=47$ days, between 1st January and 16th February 2020. An edge between a client $i$ and a server $j$ is drawn on day $t$ if the two hosts have connected at least once during the corresponding day. The number of edges ranges from a minimum of $\numprint{344565}$, observed on 1st January, to a maximum of $\numprint{912984}$ on 6th February. 

\begin{figure}[!t]
\centering
\begin{subfigure}[t]{0.475\textwidth}
\centering
\caption{Subsampling \textit{(1)}.}
\includegraphics[width=\textwidth]{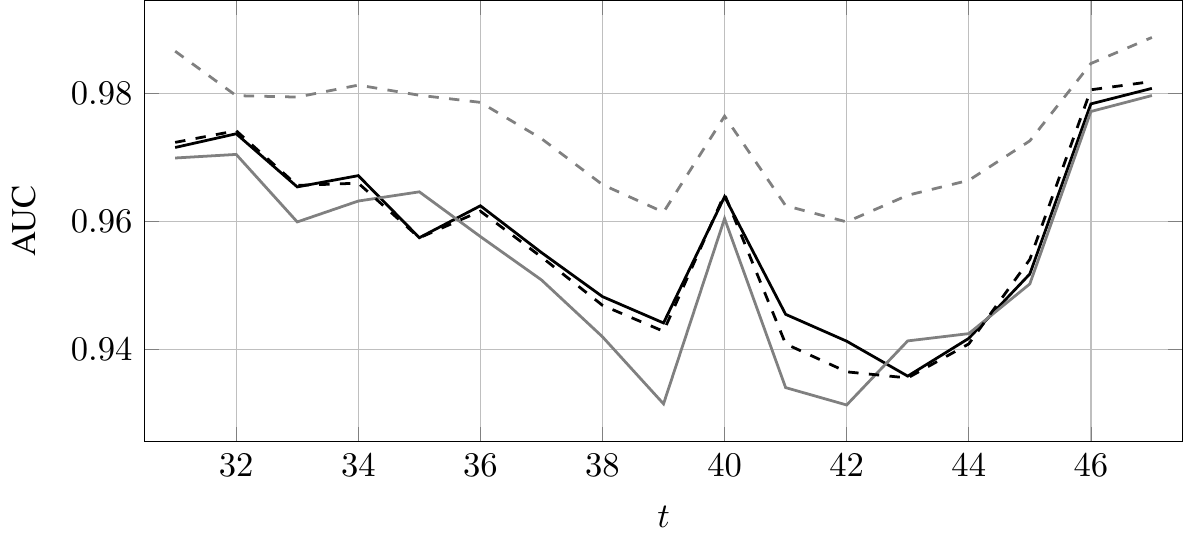}
\label{icl_results1}
\end{subfigure}
\begin{subfigure}[t]{0.475\textwidth}
\centering
\caption{Subsampling \textit{(2)}.}
\includegraphics[width=\textwidth]{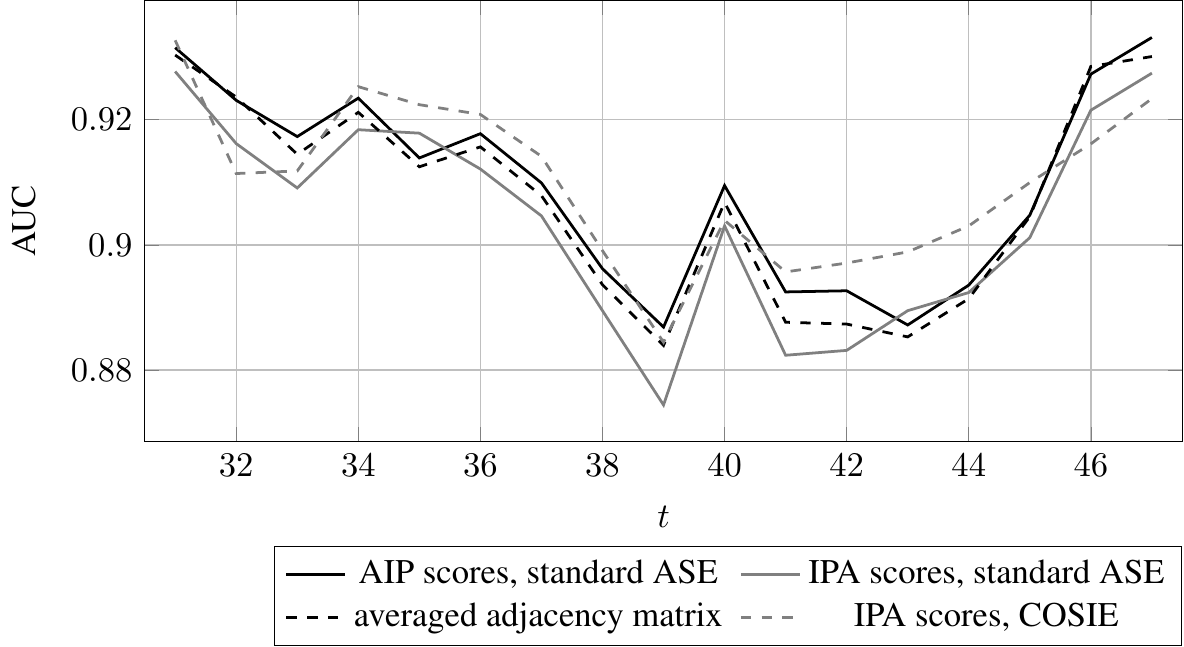}
\label{icl_results2}
\end{subfigure}
\caption{Results of the link prediction procedure on the Imperial College NetFlow network. AUCs are calculated from $6.7$ million links per graph.}
\label{icl_standard}
\end{figure}

\begin{figure}[!t]
\centering
\begin{subfigure}[t]{0.475\textwidth}
\centering
\caption{Subsampling \textit{(1)}.}
\includegraphics[width=\textwidth]{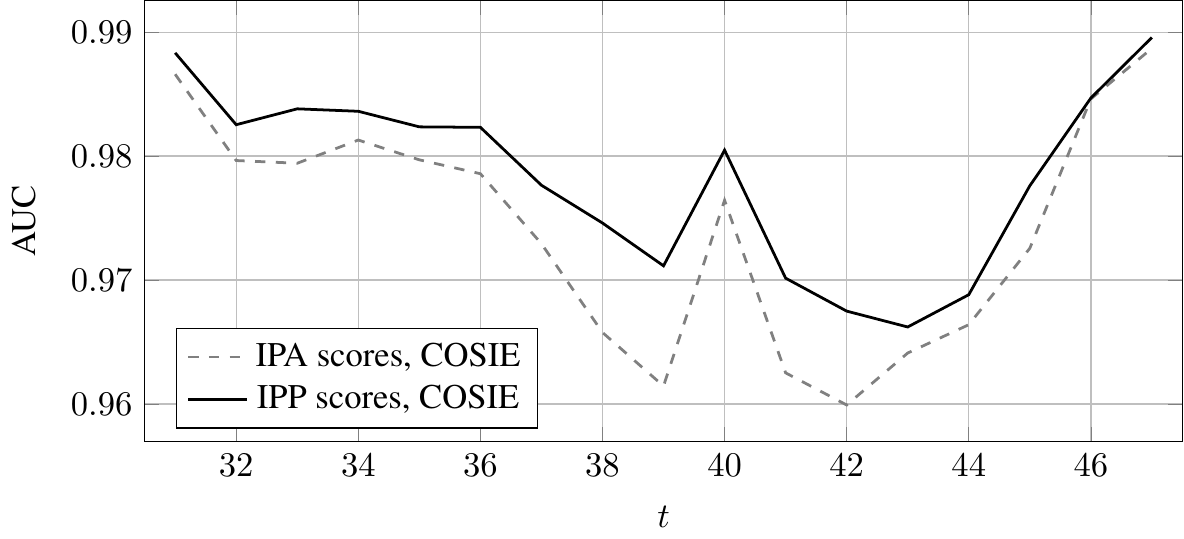}
\label{icl_cosie1}
\end{subfigure}
\begin{subfigure}[t]{0.475\textwidth}
\centering
\caption{Subsampling \textit{(2)}.}
\includegraphics[width=\textwidth]{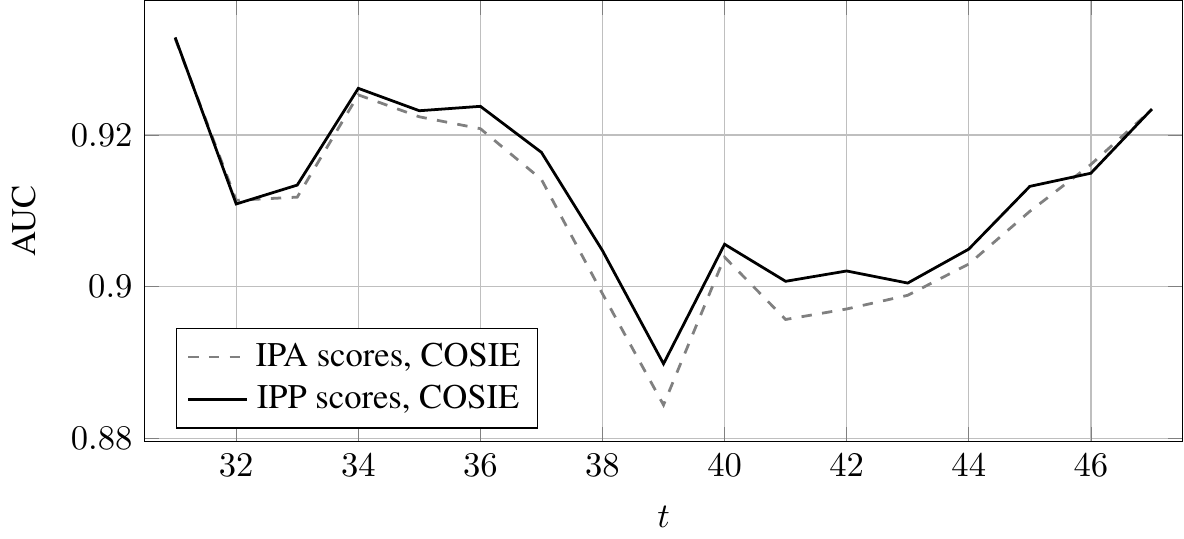}
\label{icl_cosie2}
\end{subfigure}
\caption{IPA and IPP scores using COSIE embeddings on the Imperial College NetFlow network. AUCs are calculated from $6.7$ million links per graph.}
\label{icl_cosie}
\end{figure}

The results are plotted in Figure~\ref{icl_standard}. In this case, the best performance is achieved by the COSIE scores, similarly to the LANL network in Figure~\ref{userdip_results1}. The traditional methodology of the averaged adjacency matrix is also outperformed by the AIP scores, which supports the results obtained in all the previous real data examples. As pointed out in Section~\ref{ts_extension}, fitting the time series models on the sequence $\hat{\mvec R}_1,\dots,\hat{\mvec R}_{T^\prime}$ of COSIE weighting matrices is inexpensive, and it is therefore possible to scale the time series extension of the IPA scores to a fairly large network. The results of this procedure are plotted in Figure~\ref{icl_cosie}, which shows again that the time series extension is beneficial for link prediction purposes.

\subsection{DBLP dataset} \label{dblp_section}

Finally, the methodology is evaluated on a version of the DBLP co-authorship dataset\footnote{The data are freely available at \url{https://projects.csail.mit.edu/dnd/DBLP/}.}, extensively used in the computer science literature. The DBLP network is undirected, with $n=\numprint{1258753}$ nodes, corresponding to authors of papers in computer science, for a total of $\numprint{7654336}$ undirected co-authorship edges over $T=15$ years, starting in 2000. An edge between two authors $i$ and $j$ is drawn if they co-authored a paper in year $t$. The network is extremely sparse, with edge densities ranging from $1.90\cdot10^{-7}$ in 2000 to $7.41\cdot10^{-7}$ to 2015. The number of co-authorships consistently increases over the years, corresponding to a steady increase in the number of publications in computer science journals and conferences. 

The initial $T^\prime=10$ graphs, from 2000 to 2009, are used as training set, whereas the last $T-T^\prime$ graphs, from 2010 to 2014, correspond to the test set. The results are presented in Table~\ref{dblp}, for the same models examined in Section~\ref{icl_section} on the Imperial College network, for the two different subsampling schemes. Unsurprisingly, the AIP scores with standard ASE achieve the best performance. Limited improvement is obtained considering the time series extension on the IPA scores with COSIE. 

\begin{table}[!t]
\centering
\caption{Results of the link prediction procedure on the DBLP co-authorship network. AUCs are calculated from approximately $17$ million links per graph.}
\label{dblp}
\begin{subtable}{.475\textwidth}
\centering
\caption{Subsampling \textit{(1)}.}
\label{dblp1}
\scalebox{0.75}{
\begin{tabular}{c | c c c c c}
\toprule
& \multicolumn{5}{c}{Predicted graph} \\
\multicolumn{1}{c|}{Methodology} & 2010 & 2011 & 2012 & 2013 & 2014 \\
\midrule
averaged adjacency matrix & 0.6912 & 0.6431 & 0.6095 & 0.5847 & 0.5777 \\
AIP scores, standard ASE & \textbf{0.7063} & \textbf{0.6685} & \textbf{0.6413} & \textbf{0.6235} & \textbf{0.6184} \\
IPA scores, standard ASE & 0.6637 & 0.6291 & 0.6066 & 0.5906 & 0.5856 \\
IPA scores, COSIE & 0.6744 & 0.6378 & 0.6123 & 0.5955 & 0.5903 \\
IPP scores, COSIE & 0.6893 & 0.6511 & 0.6252 & 0.6072 & 0.6020 \\
\bottomrule
\end{tabular}
}
\vspace*{1em}
\end{subtable}
\begin{subtable}{.475\textwidth}
\centering
\caption{Subsampling \textit{(2)}.}
\label{dblp2}
\scalebox{0.75}{
\begin{tabular}{c | c c c c c}
\toprule
& \multicolumn{5}{c}{Predicted graph} \\
\multicolumn{1}{c|}{Methodology} & 2010 & 2011 & 2012 & 2013 & 2014 \\
\midrule
averaged adjacency matrix & 0.6690 & 0.6202 & 0.5857 & 0.5598 & 0.5508 \\
AIP scores, standard ASE & \textbf{0.6765} & \textbf{0.6369} & \textbf{0.6083} & \textbf{0.5892} & \textbf{0.5807} \\
IPA scores, standard ASE & 0.6416 & 0.6058 & 0.5822 & 0.5653 & 0.5577 \\
IPA scores, COSIE & 0.6502 & 0.6124 & 0.5862 & 0.5687 & 0.5612 \\
IPP scores, COSIE & 0.6629 & 0.6234 & 0.5964 & 0.5773 & 0.5690 \\
\bottomrule
\end{tabular}
}
\end{subtable}
\end{table}

The performance of the RDPG on the DBLP network is significantly worse than the other examples considered in this section, suggesting that RDPG-based methods might not be the most appropriate technique for this graph. This is because the graph is extremely sparse, and a large number of new nodes appears in the network over time. Such authors, before their first published paper, are represented by rows and columns filled with zeros in the adjacency matrix, acting as disconnected components in the graph. Therefore, learning the embedding for such nodes is particularly difficult for the RDPG, since a limited (or null) history of co-authorships is available.

\subsection{Discussion and summary of results}

The main methodological contribution of this article is an adaptation for temporal link prediction of existing RDPG-based methods for joint inference on multiple graphs. 
Also, this article proposes techniques to capture the temporal dynamics of the observed graphs, combining the link probabilities and embeddings obtained via spectral methods with time series models. As demonstrated in Section~\ref{santander_alternative}, the proposed methods for combination of individual embeddings, and the extension based on time series modes, can be applied on any embedding method for static graphs, not necessarily restricted to RDPG models. The proposed methods have been applied on different synthetic and real-world datasets of different complexity, which highlighted the strengths and weaknesses of the proposed methodologies. In general, in most simulated and real-world networks analysed in this work, the AIP scores \eqref{prediction1} based on standard ASE appear to consistently achieve a very good link prediction performance in terms of AUC scores. 

Both simulations and applications on real data demonstrated that adding temporal dynamics to the estimated link probabilities via time series modelling is beneficial for link prediction purposes (\textit{cf.} Sections~\ref{sim_section}, \ref{predicted_santander_section} and \ref{lanl_predicted_scores}). 
On the other hand, the time series extension is computationally expensive for most RDPG-based models, except the IPP scores based on COSIE scores. The extensions provide significant benefits only if the network presents a strong dynamic component, as demonstrated in the simulation on the logistic dynamic network model in Section~\ref{ldnm}.

The application to the Santander bikes data (\textit{cf.} Section~\ref{santa}) highlighted that the random dot product graphs are an excellent option for link prediction for fairly small graphs without node or edge features. In such cases, the simplicity of RDPG-based models appears to overcome the advantages of deep learning methods (\textit{cf.} Section~\ref{santander_alternative}), which instead shine when applied to large graphs with rich node and edge features \citep[for example,][]{Hamilton17}. Furthermore, it must be remarked that RDPG-based methods require minimal tuning (essentially, only the choice of the latent dimension $d$), whereas alternative state-of-the-art models require computationally expensive hyperparameter tuning procedures.  

The applications on the LANL, ICL and DBLP networks (\textit{cf.} Sections~\ref{lanl_section}, ~\ref{icl_section} and~\ref{dblp_section}) demonstrated that the methodologies of AIP and IPA scores are scalable to fairly large graphs, and a good performance is achieved when the set of nodes is stable over time. It has also been shown that the proposed methodologies, in particular the AIP scores, are well suited to streaming applications (\textit{cf.} Section~\ref{stream_scores}).
The main limitation of the proposed link prediction framework is its inability to easily deal with new nodes appearing in the network, as exemplified by the application on the DBLP network in Section~\ref{dblp_section}. 

\section{Conclusion}

In this paper, link prediction techniques based on random dot product graphs have been presented, discussed and compared. In particular, link prediction methods based on sequences of embeddings, COSIE models, and omnibus embeddings have been considered. Applications on simulated and real world data have shown that one of the most common approaches used in the literature, the decomposition of a collapsed adjacency matrix $\tilde{\mvec A}$, is usually outperformed by other methods for multiple graph inference in random dot product graphs. 

Estimating link probabilities with the average of the inner products from sequences of individual embeddings of different snapshots of the graph has given the best performance in terms of AUC scores across multiple datasets. This result is particularly appealing for practical applications: calculating the individual ASEs is computationally inexpensive using algorithms for large sparse matrices, and the method seems particularly suitable for implementation in a streaming fashion, since the average inner product could be easily updated on the run when new snapshots of the graph are observed, as demonstrated in Section~\ref{lanl_section}. 

The methods discussed in the article have then been further extended to include temporal dynamics, using time series models. The extensions have shown improvements over standard random dot product graph based link prediction techniques, especially when the graph exhibits a strong dynamic component. The techniques presented in this article could also be readily extended to \textit{any} embedding method for static graphs, following the framework for calculating AIP and IPA scores presented in Section~\ref{dynamic_pred}, and the PIP and IPP extensions in Section~\ref{ts_extension}. Therefore, our methodology is not only applicable to RDPG embeddings, particularly attractive for their theoretical properties and ease of implementation, but also to modern static embedding methods for graph adjacency matrices, for example graph neural network techniques like GCN or GraphSAGE. Overall, this article provides valuable guidelines for practitioners for using random dot product graphs as tools for link prediction in networks, providing insights into the predictive capability of such statistical network models. 

\small


\bibliographystyle{rss}
\singlespacing
\bibliography{biblio}

\appendix

\section{Generalised Procrustes alignment of individual embeddings} \label{procrustes_section}

As discussed in Section~\ref{dynamic_pred}, for prediction of the future latent positions based on individual embeddings $\hat{\mvec X}_1,\dots,\hat{\mvec X}_T$, it is first necessary to align the embeddings. This section discusses a popular method for aligning two matrices: Procrustes analysis \citep{mardia} and its generalisation to $T$ matrices \citep{Gower75}. 
The alignment step is required because the latent positions $\mvec X_t$ of a single graph are not identifiable up to orthogonal transformations, which leave the inner product unchanged: for an orthogonal matrix $\bm\Omega_t$, 
$(\mvec X_t\bm\Omega_t)(\mvec X_t\bm\Omega_t)^\top=\mvec X_t\mvec X_t^\top$.
Therefore $\hat{\mvec X}_1,\dots,\hat{\mvec X}_T$ only represent estimates of a rotation of the embeddings. 
Given two embeddings $\hat{\mvec X}_1,\hat{\mvec X}_2\in\mathbb R^{n\times d}$, Procrustes analysis aims to find the optimal rotation 
of $\hat{\mvec X}_2$ on $\hat{\mvec X}_1$. The following minimisation criterion is utilised:
\begin{equation}
\min_{\bm\Omega}\left\| \hat{\mvec X}_1 - \hat{\mvec X}_2\bm\Omega \right\|_\frob, \label{proc_criterion}
\end{equation}
where $\bm\Omega$ 
is an orthogonal matrix, 
and $\|\cdot\|_\frob$ denotes the Frobenius norm. 
The solution of the minimisation problem has been derived in \cite{Schonemann1966}, and is based on the SVD decomposition 
$
\hat{\mvec X}_2^\top\hat{\mvec X}_1 = \tilde{\mvec U}\tilde{\mvec D}\tilde{\mvec V}^\top.
$
The solution is 
$
\bm\Omega^\star = \tilde{\mvec U}\tilde{\mvec V}^\top, \label{omega_star}
$ 
and it follows that the optimal rotation of $\hat{\mvec X}_2$ onto $\hat{\mvec X}_1$ is
$
\hat{\mvec X}_2\tilde{\mvec U}\tilde{\mvec V}^\top. 
\label{procrustes}
$

Similarly, a set of $T$ embeddings $\hat{\mvec X}_t\in\mathbb R^{n\times r},\ t=1,\dots,T$ can be superimposed using \textit{generalised Procrustes analysis} \citep[GPA,][]{Gower75}, which uses the minimisation criterion:
\begin{equation}
\begin{gathered}
\min_{\bm\Omega_j}
\sum_{j=1}^T \left\| \hat{\mvec X}_j\bm\Omega_j - \tilde{\mvec X} \right\|_\frob^2 \label{gpa_criterion}
\ \text{s.t.}\
\sum_{j=1}^T\mathrm{S}^2(\hat{\mvec X}_j) = \sum_{j=1}^T\mathrm{S}^2(\hat{\mvec X}_j\bm\Omega_j), 
\end{gathered}
\end{equation}
where, similarly to \eqref{proc_criterion}, $\bm\Omega_j$ 
are orthogonal matrices. Additionally, $\tilde{\mvec X}\in\mathbb R^{n\times r}$ is a reference matrix, shared across the $T$ matrices, and $\mathrm{S}(\cdot)$ is the {\it centroid size} $\mathrm{S}(\mvec M)=\|(\mvec I_n - \frac{1}{n} \mvec 1_n\mvec 1_n^\top)\mvec M\|_\frob$. The GPA algorithm solves \eqref{gpa_criterion} by iterating standard Procrustes analysis \citep{mardia}, after a suitable initialisation of the reference matrix:
\begin{enumerate}
\item {\it update} the embeddings $\hat{\mvec X}_t$, performing a standard Procrustes superimposition of each $\hat{\mvec X}_j$ on $\tilde{\mvec X}$:
\begin{equation}
\hat{\mvec X}_j \leftarrow \hat{\mvec X}_j\tilde{\mvec U}_{j}\tilde{\mvec V}_{j}^\top,
\end{equation}
where $\hat{\mvec X}_j^\top\tilde{\mvec X}=\tilde{\mvec U}_j\tilde{\mvec D}_j\tilde{\mvec V}_j^\top$;
\item {\it update} the reference embedding: $\tilde{\mvec X}=\sum_{t=1}^T \hat{\mvec X}_t$;
\item {\it repeat} steps 1 and 2 until the difference between two consecutive values of \eqref{gpa_criterion} is within a tolerance $\eta$.
\end{enumerate}
The final value of the reference embedding $\tilde{\mvec X}$ can be interpreted as the average of rotations of the initial embeddings. The alignment step 
increases the computational cost of the operations described in Section~\ref{dynamic_pred} by a factor of $\mathcal O(nd^2)$, caused by the repeated SVD decompositions and matrix multiplications in the GPA algorithm.

When $d_-\neq 0$ in the GRDPG setting, the problem is known as \textit{indefinite Procrustes problem}, and does not have closed form solution \citep{Kintzel05}. In this setting, the criterion \eqref{proc_criterion} 
must be optimised numerically
for $\bm\Omega\in\mathbb O(d_+,d_-)$, the indefinite orthogonal group with signature $(d_+,d_-)$. 
The optimisation routine could be applied iteratively in the GPA algorithm  
to obtain an \textit{indefinite GPA}. 

On the other hand, for directed and bipartite graphs, the criteria \eqref{proc_criterion} and \eqref{gpa_criterion} must be optimised jointly for the two embeddings obtained using DASE in Definition~\ref{dase}. 
For two embeddings $(\hat{\mvec X}_1, \hat{\mvec Y}_1)$ and $(\hat{\mvec X}_2,\hat{\mvec Y}_2)$ obtained from a directed graph, the solution is simply given by aligning the stacked matrices $[\hat{\mvec X}_1^\top,\hat{\mvec Y}_1^\top]^\top\in\mathbb R^{2n\times r}$ and $[\hat{\mvec X}_2^\top,\hat{\mvec Y}_2^\top]^\top\in\mathbb R^{2n\times r}$ using \eqref{proc_criterion}. 
The procedure could be also iterated for more than two embeddings to obtain a \textit{joint generalised Procrustes algorithm}.

\end{document}